\documentclass[twocolumn,twocolappendix,tighten,times]{aastex7}

\usepackage{comment}

\shorttitle{Perseus Dwarfs}
\shortauthors{Tang et al.}

\begin{document}

\title{Connection Between Dwarf Galaxies and Globular Clusters: Insights from the Perseus Cluster Using Subaru Imaging and Keck Spectroscopy}

\correspondingauthor{Yimeng Tang}
\email{ymtang@ucsc.edu}

\author[0000-0003-2876-577X]{Yimeng Tang}
\affiliation{Department of Astronomy \& Astrophysics, University of California Santa Cruz, 1156 High Street, Santa Cruz, CA 95064, USA}
\email{ymtang@ucsc.edu}

\author[0000-0003-2473-0369]{Aaron J.\ Romanowsky}
\affiliation{Department of Astronomy \& Astrophysics, University of California Santa Cruz, 1156 High Street, Santa Cruz, CA 95064, USA}
\affiliation{Department of Physics \& Astronomy, San Jos\'e State University, One Washington Square, San Jose, CA 95192, USA}
\email{aaron.romanowsky@sjsu.edu}

\author[0000-0003-1385-7591]{Song Huang}
\affiliation{Department of Astronomy, Tsinghua University, Beijing 100084, People’s Republic of China}
\email{shuang@tsinghua.edu.cn}

\author[0000-0003-2898-0728]{Nobuhiro Okabe}
\affiliation{Department of Physics, Hiroshima University, 1-3-1 Kagamiyama, HigashiHiroshima, Hiroshima 739-8526, Japan}
\email{okabe@hiroshima-u.ac.jp}

\author[0000-0002-9658-8763]{Jean P.\ Brodie}
\affiliation{Centre for Astrophysics and Supercomputing, Swinburne University, John Street, Hawthorn, VIC 3122, Australia}
\affiliation{University of California Observatories, 1156 High Street, Santa Cruz, CA 95064, USA}
\email{jbrodie@swin.edu.au}

\author[0000-0001-9742-3138]{Kevin A.\ Bundy}
\affiliation{Department of Astronomy \& Astrophysics, University of California Santa Cruz, 1156 High Street, Santa Cruz, CA 95064, USA}
\email{kbundy@ucsc.edu}

\author[0000-0003-3153-8543]{Maria Luisa Buzzo}
\affiliation{Centre for Astrophysics and Supercomputing, Swinburne University, John Street, Hawthorn, VIC 3122, Australia}
\affiliation{Astronomy Department, Yale University, 219 Prospect St, New Haven, CT 06511, USA}
\email{luisa.buzzo@gmail.com}

\author[0000-0001-6650-2853]{Timothy Carleton}
\affiliation{Arizona State University, School of Earth and Space Exploration, 781 Terrace Mall, Tempe, AZ 85287, USA}
\email{tmcarlet@asu.edu}

\author[0000-0002-6411-220X]{Anna Ferr\'e-Mateu}
\affiliation{Instituto Astrofisica de Canarias, Av. Via Lactea s/n, E-38205 La Laguna, Spain}
\affiliation{Departamento de Astrofisica, Universidad de La Laguna, E-38200 La Laguna, Tenerife, Spain}
\affiliation{Centre for Astrophysics and Supercomputing, Swinburne University, John Street, Hawthorn, VIC 3122, Australia}
\email{aferre@iac.es}

\author[0000-0001-5590-5518]{Duncan A.\ Forbes}
\affiliation{Centre for Astrophysics and Supercomputing, Swinburne University, John Street, Hawthorn, VIC 3122, Australia}
\email{duncan.forbes@gmail.com}

\author[0000-0002-2936-7805]{Jonah S.\ Gannon}
\affiliation{Centre for Astrophysics and Supercomputing, Swinburne University, John Street, Hawthorn, VIC 3122, Australia}
\email{jonah.gannon@gmail.com}

\author[0000-0003-0327-3322]{Steven R.\ Janssens}
\affiliation{Dragonfly Focused Research Organization, 150 Washington Avenue, Santa Fe, NM 87501, USA}
\email{steven.janssens@dragonfly1000.com}

\author[0009-0000-0660-1219]{Arsen Levitskiy}
\affiliation{Centre for Astrophysics and Supercomputing, Swinburne University, John Street, Hawthorn, VIC 3122, Australia}
\email{alevitskiy@swin.edu.au}

\author[0009-0008-3409-470X]{Alexi M.\ Musick}
\affiliation{Department of Physics \& Astronomy, San Jos\'e State University, One Washington Square, San Jose, CA 95192, USA}
\email{musickmalexi@gmail.com}

\begin{abstract}

We present a systematic study of 189 dwarf galaxies and their globular cluster (GC) systems in the Perseus cluster, based on deep Subaru Hyper Suprime-Cam imaging and Keck spectroscopy, supplemented by literature data. This constitutes the largest sample of dwarfs in a single galaxy cluster to date with simultaneous deep imaging, spectroscopic coverage, and GC measurements, while uniquely spanning a broad and continuous range of galaxy properties. 
We find an anti-correlation between GC specific mass and galaxy stellar mass for dwarfs in Perseus similar to observations in other clusters. At fixed stellar mass, dwarfs with lower surface brightness or larger effective radius tend to be more GC-rich -- suggesting either high GC formation efficiency in an earlier compact-galaxy phase, or less efficient GC disruption.
The correlation between GC richness and axis ratio in Perseus is weaker than in other environments. We find some connection between GC richness and infall time, but not with the clear correlations found in Virgo, Coma, and cosmological simulations. More complete observations are needed to test for cluster-to-cluster variations in galaxy and GC evolutionary histories.
This work demonstrates the potential of new wide-field imaging and spectroscopy surveys for understanding GCs and dwarf galaxies, and highlights the need for further work in theoretical modeling.

\end{abstract}

%% The AAS Journals now uses Unified Astronomy Thesaurus concepts:
%% https://astrothesaurus.org
\section{Introduction} \label{sec:intro}

Globular clusters (GCs) are old, massive, and dense star clusters with stellar masses ranging from $10^4$ to $10^6\ M_{\odot}$. They are thought to originate in some of the earliest and most intense episodes of star formation, and thus represent important probes of primordial epochs \citep{Brodie2006}. 

Decades of studies from the ground and space have established basic scaling relations between GC systems and their host galaxies. Notably, the fraction of stellar mass in GCs exhibits a minimum for galaxies near Milky Way mass, and increases dramatically for lower and higher masses (e.g., \citealt{Peng2008,Liu2019}). This trend is a mirror image of the stellar-to-halo mass relation, suggesting that the ratio of GC system mass to halo mass is remarkably constant on average (e.g., \citealt{Spitler2009,Georgiev2010,Hudson2014}), with a value of $\sim 0.003$\% \citep{Harris2017}.

Although dwarf galaxies tend to have higher GC to galaxy stellar mass fractions on average, there is a large galaxy-to-galaxy scatter, which is conventionally attributed to environmental effects. Dwarf galaxies currently found in the central regions of groups and clusters appear more GC-rich \citep{Peng2008,Lim2018}. These galaxies are thought to have formed early, with dense gas and bursty star formation more conducive to GC formation \citep{Mistani2016,Ramos-Almendares2020}. They also quenched early, thereby reducing the total stellar mass \citep{Carlsten2022}. In contrast, dwarfs in lower-density regions had more protracted and gentler histories, with less GC formation and more field star formation \citep{Garrison-Kimmel2019}. However, the support from observations for this picture is surprisingly incomplete, consisting mainly of correlations in small samples of cluster dwarfs, along with very sparse studies of GCs in field and group dwarfs.

A special type of low-mass galaxies, known as ultra-diffuse galaxies (UDGs; \citealt{vanDokkum2015}), has recently attracted growing attention and discussion. UDGs are characterized by their unusually large effective radii for dwarf masses ($R_{\rm e}>1.5$ kpc) and very low surface brightness (SB; $\mu_{g,0}>24\ {\rm mag\ arcsec^{-2}}$). From a morphological and structural perspective, UDGs appear to be a natural extension of classical dwarf galaxies in parameter space \citep{Amorisco2016,Conselice2018,Zoller2024}. However, many of them host distinctly more populous GC systems compared to other dwarfs at similar masses \citep{vanDokkum2017,Lim2018,Forbes2020,Danieli2022,Saifollahi2025a}, suggesting different pathways in their formation mechanisms. Therefore, understanding their origins needs incorporating additional properties into the discussion.

Conventionally, GC-rich UDGs are interpreted as earliest-infall dwarfs, whose large sizes reflect prolonged interaction with the cluster environment \citep{Carleton2021, FerreMateu2023}, positioning them as a tail of the normal dwarf population. In this case, UDGs that infall later have more extended star formation histories, and evolve into more disky configurations with less GC formation \citep{Lim2018,Pfeffer2024}. However, the discovery of GC-rich UDGs in late-infall or field environments has challenged this view (e.g., \citealt{Gannon2022,Janssens2022, Forbes2023}). Some GC-rich UDGs are more consistent with a ``failed galaxy'' scenario, where intense early starbursts followed by rapid quenching produce high GC abundance, low stellar mass, and large size (e.g., \citealt{Forbes2020,Saifollahi2022,Toloba2023,FerreMateu2023,Buzzo2024}).

The diversity of UDGs suggests that the scatter in GC mass fraction among low-mass galaxies may also fundamentally correlate with other properties beyond environment in galaxy clusters, including morphology and stellar populations. However, except for a few studies (e.g., \citealt{Prole2019}), analyses of GCs in cluster dwarfs so far have mostly focused either on the high-SB end or on UDGs, leaving gaps in parameter space without including dwarfs of intermediate mass and/or SB (e.g., \citealt{Lim2018,Lim2020}). Furthermore, previous work often relied on projected clustercentric radius as the primary proxy for local environment -- only partially capturing the true environmental influences on dwarf galaxies. A more robust approach involves phase-space analysis (i.e., position–velocity distribution), which provides a clearer indication of the infall stage and environmental history (e.g., \citealt{Rhee2017}), and has been applied in recent studies of UDGs \citep{Alabi2018,Gannon2022,FerreMateu2023,Forbes2023}. Intriguingly, those results so far showed no clear connection between GC richness and infall times, supporting diverse origins of UDGs. However, the focus has still been essentially on UDGs in this space, lacking connection to average dwarfs.

Considering that individual clusters and their constituent galaxies may vary in their assembly histories and properties, a homogeneous analysis of a large, continuous sample of low-mass galaxies in parameter space within the same galaxy cluster is essential. While classical dwarfs show no significant dependence of GC richness on galaxy cluster mass \citep{Lim2018,Liu2019}, GC-rich UDG populations appear to be predominantly found in massive clusters like Coma but are largely absent in lower-mass clusters or galaxy groups. In addition to Coma, the Perseus cluster (at 75 Mpc, systemic velocity $V_0=5258$ km s${}^{-1}$, velocity dispersion $\sigma=1040$ km s${}^{-1}$, virial radius $R_{200}=2.2$ Mpc, virial mass $M_{200}=1.2\times10^{15}\ M_{\odot}$, \citealt{Aguerri2020}) is another high-mass galaxy cluster in the nearby Universe that serves as an ideal laboratory for GC-related studies. Many studies have already included or focused on dwarf galaxies and UDGs in the Perseus cluster (e.g., \citealt{Brunzendorf1999,Conselice2003,Penny2011,Penny2014,Wittmann2017,Wittmann2019,Gannon2022,Janssens2024,Li2025}). Taking advantage of Euclid Early Release observations (ERO), \cite{Marleau2025} and \cite{Saifollahi2025} investigated the connection between GCs and the intrinsic properties of Perseus dwarfs based on a large sample. While their findings were broadly consistent with previous studies, there was little focus on the role of environmental effects in shaping GC systems. Furthermore, velocity measurements are unavailable for the vast majority of the dwarfs in this sample, limiting phase-space analyses.

In this work, we conduct a differential study of dwarf galaxies in the Perseus cluster, with a wide range of infall times, GC richness, and morphologies. Compared with previous studies, we are able to take advantage of a larger dataset, homogeneous analysis, and a combination of photometry from the Subaru Hyper Suprime-Cam (HSC; \citealt{Miyazaki2018,Aihara2018}) and spectroscopy from the Keck II telescope.

We structure this paper as follows. In Section \ref{sec:data}, we describe our Keck observations and the Subaru HSC imaging used in this work. We present our data analysis methods in Section \ref{sec:methods}. The results and discussions are shown in Sections \ref{sec:results} and \ref{sec:discussion}. A summary follows in Section \ref{sec:conclusion}. All magnitudes mentioned in this work are in the AB system. The Galactic extinction for the Perseus Cluster is significant, with the correction near cluster center of 0.52, 0.38 and 0.28 mag in $g$, $r$, $i$ filters. The correction for each galaxy varies slightly depending on their locations on the sky. Unless otherwise noted, all the magnitudes and colors mentioned below are corrected for Galactic extinction.

\section{Data} \label{sec:data}

\subsection{Subaru Hyper Suprime-Cam imaging}

We use archival HSC imaging of the Perseus Cluster in the $g$, $r$, and $i$ bands, which were observed in 2014 September and November. The total exposure times for the three filters are 23760s, 4320s, and 2160s, respectively. Here the $g$-band imaging is much deeper than used in our previous work on Perseus \citep{Gannon2022}, and is the same as in the weak lensing study of \citet{HyeongHan2025}. There is also $z$ band imaging available, but the signal-to-noise ratio (S/N) is relatively low and we do not use it for this project. The imaging covers a circular region of 1.5-degree diameter, reaching out to $\sim 0.5\ R_{200}$.

We reduced the HSC images using the standard pipeline with version 8 \citep[hscPipe8;][]{Bosch2018}, including overscan, bias and dark subtraction, and flat fielding. We note that the sky subtraction from this pipeline version is not optimized for LSB features, resulting in modest systematic uncertainties in the galaxy photometry compared to the shallower imaging from \citet{Gannon2022}, as will be discussed later in Section \ref{sec:galfit}. However, we chose to use the deeper data for better results on the GCs. We used the Pan-STARRS photometric catalog \citep{Magnier2020} for astrometry and zero-point calibration of each CCD. We combined multiple images, taking into account the sky correction pattern. The average seeing in the coadded images is 0.69, 0.52, and 0.62 arcsec, respectively. With this exquisite image quality, the 5$\sigma$ depth for point sources in each band is about 28.2, 26.5, and 26.0 mag, respectively -- allowing for relatively complete detections of GCs in Perseus (Section~\ref{sec:gc_richness}).

\subsection{Dwarf galaxy sample}

\begin{figure*}
\centering
\includegraphics[width=0.9\textwidth]{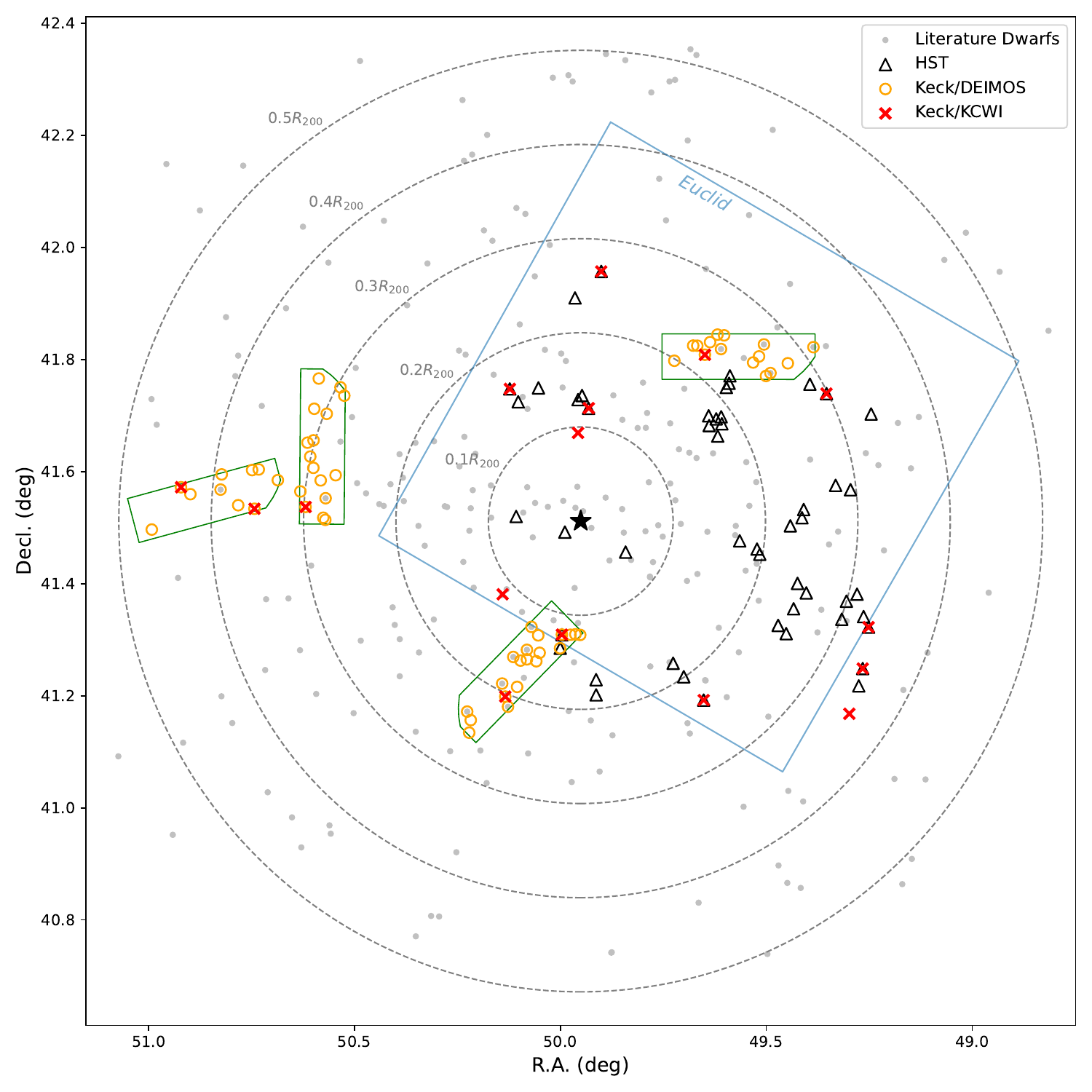}
\caption{Sky positions of dwarf galaxies in the Perseus cluster. Those observed by Keck/DEIMOS are marked with orange open circles (showing only the confirmed cluster members), by Keck/KCWI with red crosses, and by HST with black open triangles. The four DEIMOS footprints are overplotted as large green polygons. The cluster center galaxy NGC 1275 is marked as the black star, and the gray dashed concentric circles indicate clustercentric radii from 0.1 to 0.5 $R_{200}$. The light gray dots represent the member dwarfs in this sky region confirmed with spectroscopy in the literature \citep{Kang2024}. The footprint of the Euclid observation is also overplotted.}
\label{fig:dwarf_spatial_distribution}
\end{figure*}

We establish our sample of 190 dwarf galaxies to be studied in this project by following the steps outlined below, noting that the sample evolved during the course of the project based on data availability.

Our imaging data sources for LSB galaxy target selection are discussed in \cite{Gannon2022}, based on identifications in CFHT/MegaCam imaging (`R' designations), and in deeper HSC imaging (`S' and `H' designations). Around 700 candidates were found in HSC using SExtractor \citep{Bertin1996}, including reidentifications of many objects from MegaCam. For this project, we initially
%visually inspected all of these galaxies, and 
classified all of these galaxies as GC-rich, GC-mid, and GC-poor, based on a visual assessment of the fraction of galaxy light in GCs. 
A similarly qualitative approach was found in \cite{Gannon2022} to yield results consistent with high-quality quantitative results from HST imaging. Our project design is oriented around a differential study of GC-richness, and therefore we focused on large LSB dwarfs (mostly UDGs) that were clearly GC-rich or GC-poor
(which will be quantified more rigorously below).
Based on the field of view ($16' \times 4'$) of the DEep Imaging Multi-Object Spectrograph (DEIMOS; \citealt{Faber2003}) on the Keck II telescope, we selected regions of Perseus that include as many of the high-priority UDGs as possible within a DEIMOS footprint.

Then, working with the design constraints of DEIMOS slitmasks, we added potential Perseus cluster dwarf galaxies within the DEIMOS field of view as fillers. Finally, 93 targets were observed in 4 different pointings, and 71 of them are confirmed to be Perseus members. The details of the DEIMOS observations and data are presented in Section \ref{sec:deimos}. 12 dwarfs with relatively high SB were also identified as Perseus members in \cite{Kang2024}.

Observations with the Keck Cosmic Web Imager (KCWI; \citealt{Morrissey2018}) are available for 16 Perseus dwarfs and UDGs (see Section~\ref{sec:kcwi}), with six of them also included among our DEIMOS targets. Therefore, the additional 10 galaxies will also be incorporated into our sample.

50 LSB dwarfs have been observed by the Hubble Space Telescope (HST), with their GC systems studied \citep{Janssens2024,Li2025}\footnote{We do not include the extreme cases CDG-1 and CDG-2 since the host galaxy light is poorly constrained or undetected \citep{Li2022,Li2025b}}. Nine of them are among our Keck targets. Although spectroscopy is not available for the remaining 41 dwarfs to determine their Perseus membership, we include them in our sample because the other nine dwarfs are all confirmed to be associated with the cluster. We omit only one galaxy (W8) as it is too faint for reliable HSC photometry. We will use this HST subsample only for the parts of the analysis which do not require recessional velocity information.

\begin{figure*}
\centering
\includegraphics[width=0.75\textwidth]{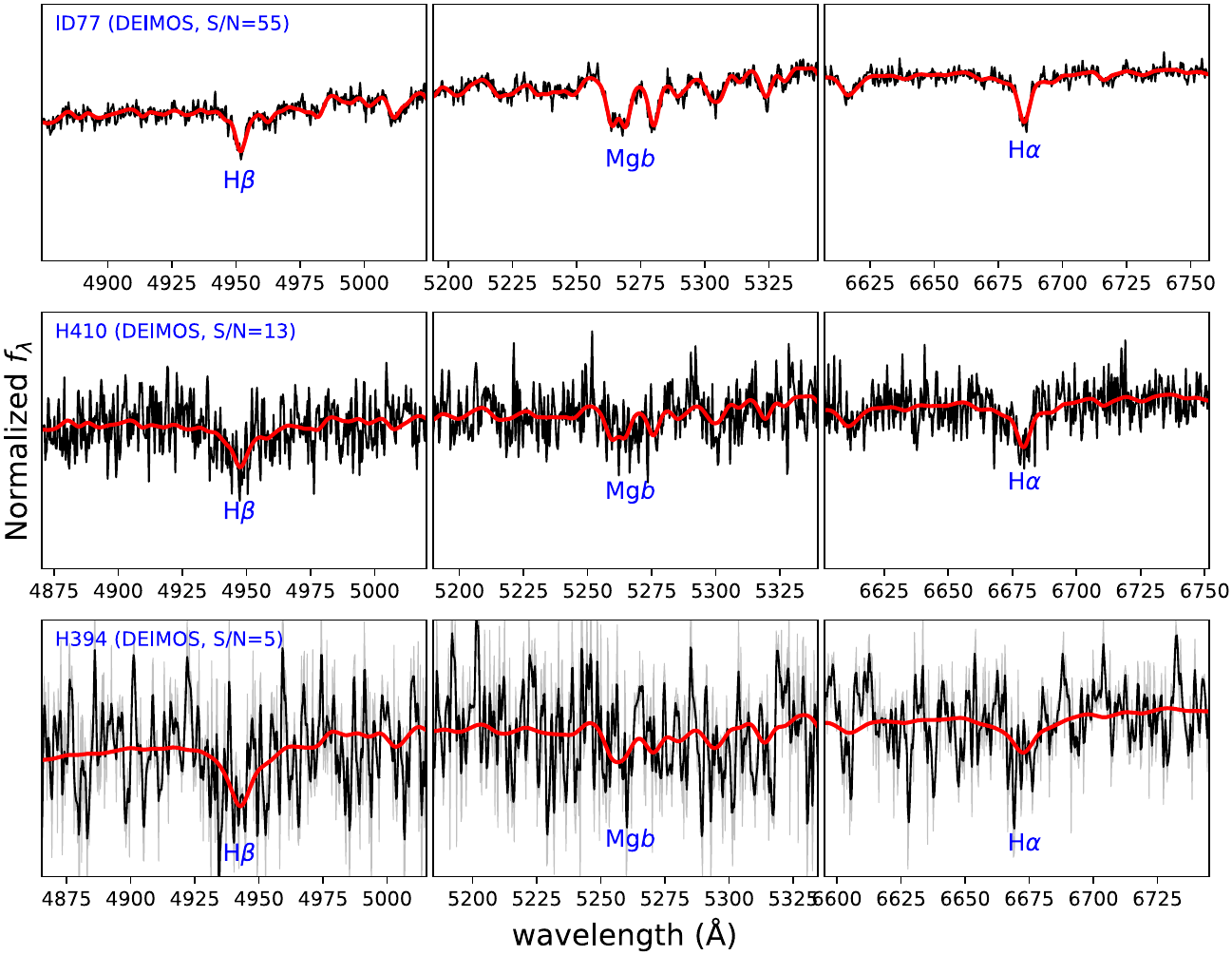}
\caption{Example pPXF fits of DEIMOS spectra with different continuum S/N. From top to bottom, these three galaxies are selected from the brightest, medium, and faintest thirds in our sample of dwarfs with Keck observations. From left to right, the three wavelength windows show the fitting of the major absorption features, including H$\beta$, Mg$b$, and H$\alpha$. The observed spectra are shown in black except for the bottom row of low S/N, where the observed spectrum is colored gray, and black represents the smoothed spectrum for more reliable fits. The best-fit model is shown in red. These major absorption features can be well fitted with medium to high S/N. Under low S/N conditions, at least H$\alpha$ and/or H$\beta$ can be identified for a sufficiently accurate recession velocity estimate. We note that all of our KCWI spectra have medium to high S/N ($\gtrsim 10$).}
\label{fig:ppxf_example}
\end{figure*}

\cite{Marleau2025} have obtained a much larger dwarf galaxy sample in Perseus from Euclid imaging. However, since up to 20\% of these dwarfs may not be Perseus members, we prefer not to directly merge their sample with our sample in this study. \cite{Kang2024} provided a recent catalog of spectroscopically confirmed Perseus member galaxies, combining literature and new redshifts\footnote{Another large dataset of Perseus redshifts was obtained by \citet{Aguerri2020} but would add relatively few galaxies to our sample.},
down to $r$-band magnitudes of $\sim$19, entering the mass range of dwarfs. By cross-matching the \cite{Marleau2025} dwarf sample with the \cite{Kang2024} catalog, we obtain an additional 78 Perseus dwarfs without Keck observation. Compared to the 71 galaxies selected above with Keck data, these 78 galaxies generally have higher SB or are nucleated, making them more easily detectable in the shallower spectroscopic data in \cite{Kang2024}. The GC measurements by \cite{Marleau2025} and \cite{Saifollahi2025} for the dwarfs in the Euclid footprint are highly valuable for our study. In the absence of HST observations in our sample, we prioritize using the Euclid GC results which are well consistent with our measurements from HSC and HST (see Section \ref{sec:gc_richness}).

The final sample consists of 189 dwarf galaxies, with 71 of them having spectroscopic observations from Keck, 91 of them having known recessional velocities from \cite{Kang2024}, 49 of them with HST observations, and 144 with Euclid observations. Their sky locations are shown in Figure \ref{fig:dwarf_spatial_distribution}, with an image gallery presented in Figure \ref{fig:gallery}. Some of the LSB dwarfs in this sample (e.g., R15, R16, R21, R24, R84, S74) have also been analyzed in previous papers on dynamics and stellar populations (e.g., \citealt{Gannon2022,Buzzo2022,FerreMateu2023,FerreMateu2025,Forbes2025,Levitskiy2025}).
As described above, our sample comprises various subsets, each potentially biased in SB, mass, or other properties. Interestingly, however, the combined sample exhibits a broad and uniform distribution across parameter space, showing no obvious bias (see Section \ref{sec:host}). We do not use any explicit criteria to exclude star-forming dwarfs, but almost all of our dwarfs are quiescent, which is expected in the Perseus cluster environment. Our sample spans a stellar mass range of of $7.0 \lesssim \log(M_*/M_\odot) \lesssim 9.5$, magnitude range of $22.5 \gtrsim m_g \gtrsim 16.5$, and SB of $26.5 \gtrsim \langle \mu_{g} \rangle_{\rm e} \gtrsim 20.5$ mag arcsec$^{-2}$.

\subsection{Keck/DEIMOS observation}
\label{sec:deimos}

The spectroscopic data used in this work were mainly obtained by Keck/DEIMOS in program U116 (PI: J.~Brodie). The dates of our observations are 2022 October 31, November 1, 23 and 24. The weather and sky transparency were good, and the seeing varied from $0.8''$ to $1''$. The 1200B grating with the GG400 filter was used, covering a wavelength range of $\sim$ 4500--7500~\AA\ with a spectral resolution of $R \sim$ 1000. The slit width was set to $3''$ to collect more photons from the LSB galaxies and match the typical size of the Perseus UDGs. This set-up is similar to that of the Coma cluster observations in \cite{Alabi2018} and \cite{FerreMateu2018}, except for using a different grating. We designed four slit masks to cover four different regions of the Perseus Cluster within $\sim 0.5 R_{200}$ (shown in Figure \ref{fig:dwarf_spatial_distribution}), with exposure times of 4.25h (1800s$\times$8$+$900s), 7.00h (1800s$\times$14), 6.50h (1800s$\times$13), and 3.83h (1800s$\times$7$+$1200s) from East to West, respectively.

We use the Python-based semi-automated reduction pipeline PypeIt \citep{Prochaska2020} to reduce our DEIMOS data. Under the default settings, PypeIt performs bias subtraction, flat fielding, edge tracing, and wavelength calibration for each individual science exposure, and then co-adds the data for each of the four masks. For each slit, we perform optimal extraction and subtract the local sky. The initial guess for the extraction width is set relatively large to avoid shredding. Finally we obtained spectra of 93 galaxies with DEIMOS.

\begin{figure*}
\centering
\includegraphics[width=0.7\textwidth]{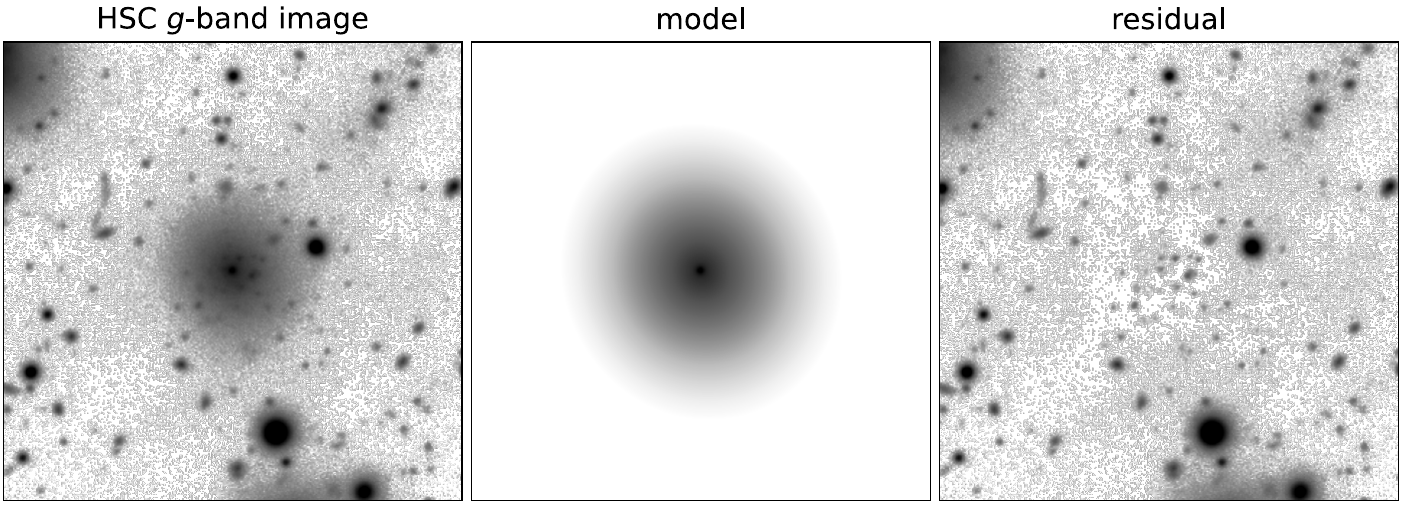}
\caption{An example of our photometric modeling of HSC imaging using GALFIT, for the LSB dwarf galaxy R41. The three panels show the original image, model, and residual from left to right. The images have a side length of 10 times the $g$-band major-axis $R_{\rm e}$, which is the default fitting box for all our galaxies.}
\label{fig:galfit_example}
\end{figure*}

\subsection{Keck/KCWI observations}\label{sec:kcwi}

In addition to DEIMOS, spectroscopic observations with Keck/KCWI provide more LSB dwarfs in the Perseus cluster for study. The data for 16 dwarfs (mostly UDGs) were collected in 19 nights in the fall from 2018 to 2024. These dwarfs have been studied elsewhere (\citealt{Gannon2022,FerreMateu2023,Levitskiy2025}; Levitskiy et al. in prep.), with well-measured radial velocities provided, so we will not repeat the practice here. The only exception is PMDG-47, which is reported exclusively here, and follows a similar process to that described in \cite{Gannon2020}.

\section{Methods} \label{sec:methods}

\subsection{Spectral fitting} %\label{sec:velocity}

To confirm membership in the Perseus cluster of our target galaxies, and to further determine their infall stages, we first use Penalized Pixel Fitting (pPXF; \citealt{Cappellari2004}) to perform full spectral fitting and obtain their recessional velocities. For the spectral templates, we use MILES single stellar population models \citep{Vazdekis2016}, with the Padova+00 isochrones \citep{Girardi2000} and \cite{Kroupa2001} initial mass function. The free kinematic parameters in the fitting are mean velocity and velocity dispersion, without any higher-order Gauss--Hermite moments included. The orders of the additive and multiplicative polynomials are set as 15 and 5, respectively, to sufficiently adjust the continuum shape of the templates to match our observed spectra. In Figure \ref{fig:ppxf_example}, we show example fits with a wide range of continuum S/N. Even at S/N as low as $\sim5$, we are still able to rely on the major absorption features such as H$\alpha$ and/or H$\beta$ to fit the recessional velocity. The measured velocities are listed in Table \ref{tab:data_properties}. We do not see strong emission lines in any of our dwarfs.

Six galaxies in our sample were observed with both DEIMOS and KCWI, providing us with a rough uncertainty estimate of our line-of-sight velocity measurements. We find that the velocity difference when comparing different data is $\lesssim 50$ km s${}^{-1}$, which is negligible for phase-space analysis. For the faintest dwarfs in our sample, uncertainties can reach up to $\sim 100$ km s${}^{-1}$, but this still does not significantly affect the determination of their Perseus cluster membership or infall stage.

Since our sample is selected within the projected distance of $0.5 R_{200}$ from the Perseus center (limited by the HSC field of view), we find 71 dwarf galaxies with velocity differences from the mean velocity of the cluster ($V_0=$ 5258 km s${}^{-1}$) of no more than $\sim$ 2.5 times the cluster velocity dispersion ($\sigma=$ 1040 km s${}^{-1}$), which we classify as Perseus members (the same criterion used in \citealt{Gannon2022}). Most of the rejected objects have redshifts greater than 0.1. Later in this paper, the infall stages of these dwarfs will be further determined in phase space.

\subsection{Photometric modeling of galaxy light}
\label{sec:galfit}

We use GALFIT \citep{Peng2002} to perform two-dimensional galaxy modeling to derive the morphological properties and magnitudes of the 190 dwarfs in our sample. The procedure is basically the same as described in \cite{Tang2025a}. A combination model is chosen for fitting, consisting of a single S\'ersic model for the galaxy light and a plane sky model for the background, which provides reasonable fits to all of the images. When a galaxy is nucleated, an additional PSF component is included at the center. We run SExtractor with a low detection threshold to generate masks, aiming to minimize contamination from foreground stars and background galaxies. The morphological parameters are fitted in each filter independently to allow for variation with wavelength, but the galaxy center is fixed to the $g$-band result. The cutout images used for fitting have widths equal to 10 times the $g$-band major-axis $R_{\rm e}$ of each galaxy. We run GALFIT iteratively, updating the initial parameter guesses, as well as the cutout sizes and masks in each iteration, until the results converge.

As an example of the fits, we show the GALFIT best-fitting model and residual for the $g$ band in Figure \ref{fig:galfit_example}. Table \ref{tab:data_properties} and \ref{tab:data_properties_euclid} summarize the properties and photometric results for each galaxy in our sample.

We caution that for larger galaxies which may be affected more by the imperfect sky subtraction of the HSC imaging, the magnitudes and $R_{\rm e}$ presented in Table \ref{tab:data_properties} and \ref{tab:data_properties_euclid} may not be particularly accurate (see discussion in Appendix~\ref{sec:dwarf_gallery_properties}). However, this inaccuracy does not have any significant impact on the results and conclusions of this work. 

There are 40 UDGs in our sample in line with the UDG definition of the average surface brightness $\langle \mu_{g} \rangle_{\rm e} \gtrsim 25\ {\rm mag\ arcsec^{-2}}$ and size $R_{\rm e}>1.5$ kpc (e.g., \citealt{Yagi2016,vanderBurg2017,Janssens2019,Gannon2022}), 20 of which have recessional velocity measurements.

\begin{figure*}
\centering
\includegraphics[width=0.95\textwidth]{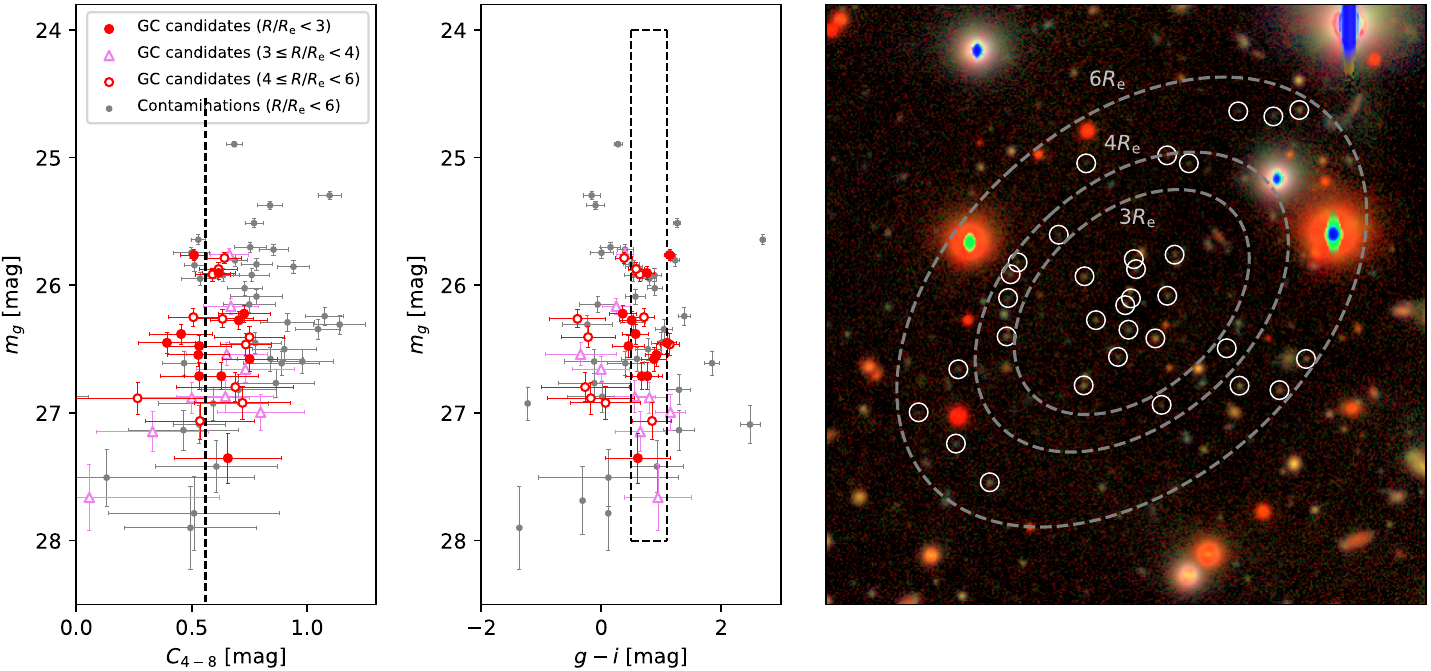}
\caption{An example demonstrating our GC selection criteria, for UDG W89. {\it Left and Middle panels}: distribution of the concentration parameter $C_{4-8}$ and the $g-i$ colors, versus $g$-band magnitudes of the detected sources. The errorbars represent 1-$\sigma$ uncertainties of each property. The red solid and open circles represent GC candidates within 3 $R_{\rm e}$ and between 4 and 6 $R_{\rm e}$, respectively, and the violet triangles are GC candidates between 3 and 4 $R_{\rm e}$. These GC candidates agree with the point source $C_{4-8}$ (vertical dashed line in the left panel) and GC color and magnitude range (dashed box in the middle panel) within 2$\sigma$. The sources that do not meet these criteria are considered as contaminants, shown as the gray points. {\it Right panel}: spatial distribution of the selected GC candidates within 6 $R_{\rm e}$, the same as the red circles and violet triangles in the left two panels. The galaxy diffuse light has been subtracted from the image. The three dashed ellipses correspond to radii of 3, 4, and 6 $R_{\rm e}$.}
\label{fig:gc_selection_example}
\end{figure*}

\subsection{GC richness}
\label{sec:gc_richness}

Ground-based telescopes like Subaru are not comparable to space-based telescopes such as HST and Euclid in terms of counting individual GCs at the Perseus distance. However, the brighter GCs are relatively easy to identify and, due to the roughly lognormal shape of the GC luminosity (or mass) function \citep[e.g.,][]{Harris2001,Brodie2006}, they contribute the majority of the total GC system mass. In this sense, GC system mass is less sensitive to the faint end of the GC population, which is difficult to distinguish from contamination. Given the small scatter in the average GC mass for dwarf galaxies \citep{Harris2017}, GC number and total mass are expected to correlate tightly, although recent studies have identified exceptions with top-heavy GC luminosity functions \citep[e.g.,][]{Shen2021, Janssens2022, Tang2025b, Li2025}. Therefore, in this work, we adopt the mass fraction in the GC system relative to the galaxy stellar mass as our indicator of GC richness, rather than the specific frequency more often used in the literature based on GC number \citep[e.g.,][]{Peng2008,Lim2018}.

We follow a traditional way of selecting GC candidates using color and concentration, keeping in mind that this work at such a large distance has rarely been attempted from the ground, and requires deep imaging with excellent seeing (e.g., \citealt{vanDokkum2016}). To determine a reasonable color range of GC candidates, we generate predictions from single stellar population models with ages of 3--14 Gyr and metallicities [M/H] from $-2.0$ to 0 dex, based on MILES models. We adopt the color range $0.5<(g-i)_0<1.1$ mag from these models for GC selection. Given the resolution of HSC, GCs appear as point sources at the Perseus distance.

In preparation for GC analysis, we begin with the GALFIT residual images and further clean them by subtracting a smooth background obtained by median filtering. We find that the stacked $g+r$ image is the best for source detection, as it leverages the depth of the $g$-band and the high spatial resolution of the $r$-band.

With the two key GC selection criteria determined (i.e., color and concentration), we run the DAOPHOT algorithm \citep{Stetson1987} on the fully cleaned $g+r$ image to generate source catalogs, with a detection level of 3$\sigma$. We find DAOPHOT works better here for point source detection than SExtractor, but we also use SExtractor to generate a segmentation map to mask bright contaminants. For the detected sources, we do aperture photometry with apertures of 4 and 8 pixels in diameter. The concentration parameter $C_{4-8}$ is therefore defined by the magnitude difference between the two apertures on the stacked $g+r$ image, as a discriminant between point-like sources and extended objects. We require the selected GC candidates to be consistent with the $C_{4-8}$ of point sources within 2$\sigma$ (measurement uncertainty), which is determined by the nearby stars of each galaxy since the PSF has a small variation across the field. In other words, there is a magnitude dependence on the concentration selection.

To get the total magnitudes for each source, we make an aperture correction for our 4-pixel diameter photometry. The color is calculated based on the aperture-corrected magnitudes, and the GC candidates need to fit with the color range $0.5<(g-i)_0<1.1$ within 2$\sigma$. 

A further magnitude cut $24<m_g<28$ (or $5 \times 10^4 \lesssim M_\star/M_{\odot} \lesssim 2\times 10^6$, assuming a mass-to-light ratio of $M/L_g = 1.4$) is applied in $g$-band. We find this $M/L_g$ generally corresponds to the typical $M/L_V \sim 1.3$ for low-mass galaxies \citep{Harris2017}, by modeling simple stellar populations in MILES. Given the typical GC mass of $\sim 10^5 M_{\odot}$ for dwarfs \citep{Harris2017} and the lognormal shape of the GCLF, the relatively few low-mass GCs missed by this cut do not contribute significantly to the total GC mass estimate (which will be our quantity of interest, rather than GC number). Therefore, these fainter objects are excluded for simplicity. On the other side, the brighter limit aims to avoid including nuclei and ultracompact dwarfs. Figure \ref{fig:gc_selection_example} shows an example of our GC selection process.

In order to estimate the completeness, we adopt an artificial point source test. We randomly inject 20000 artificial stars (using the PSF generated by Photutils) into the blank fields of our $g+r$ detection image. The artificial stars are placed in batches of 2000, with a uniform $g+r$ magnitude distribution. This procedure does not really account for the crowding of the GCs themselves toward the centers of the galaxies, potentially losing even bright GCs from blending that produces a high concentration index. However, this appears to be rare in practice for the LSB dwarfs. We model the completeness using Eq. (2) in \cite{Harris2016}, and find the 50\% completeness limit corresponds to $m_{g,0}\sim28.2$ if assuming a typical $g-r$ color of $\sim0.6$ for GCs in dwarf galaxies. Note that this artificial point source test is performed over the entire HSC field of view. While the noise level is relatively uniform on large scales, in practice, fluctuations of $\sim$15\% can be observed across the entire image. Such fluctuations may introduce a variation of about 0.15 mag to the completeness limit, and there are also extinction variations of $\pm 0.1$~mag. However, considering the expected GCLF turnover magnitude of $m_g \sim 26.8$ mag at the Perseus distance (absolute turnover magnitude $M_g \sim -7.5$; \citealt{Villegas2010}), the completeness limit is already more than one magnitude deeper. We therefore find the completeness correction is almost negligible and does not need to be applied here. The dominant source of uncertainty is the contamination correction.

To calculate the total GC mass $M_{\rm GC}$ for each dwarf, we start with the GC candidates within an ellipse region extending to 3$\times$ the major-axis $R_{\rm e}$ from the galaxy center, and then use the other candidates in the elliptical annulus between 4 and 6 $R_{\rm e}$ to estimate the ``background'' contribution (which can include contributions from foreground stars, background galaxies, and intracluster GCs). \cite{Janssens2024} and \cite{Saifollahi2025} showed that for low SB dwarfs in Perseus, the typical ratio between $R_{\rm GC}$ (i.e., half-number radius of GC system) and $R_{\rm e}$ is about 1.5. If the GCs follow an exponential distribution with this $R_{\rm GC}/R_{\rm e}$ ratio, only $\sim$15\% of the GCs would lie beyond 3 $R_{\rm e}$. We adopt a stellar mass-to-light ratio of 1.4 in $g$-band as mentioned above to estimate the stellar mass of each GC candidate. 
For the GC candidates within the magnitude range above, we calculate the ``mass excess'' within 3 $R_{\rm e}$ above the background level. We find that background estimates derived for individual galaxies are strongly affected by stochastic fluctuations. Therefore, we adopt a unified background estimate given by the mean background density measured across all our galaxies from their 4--6 $R_{\rm e}$ annuli, which is $\sim 1.6 \times 10^4 \ M_{\odot}\ {\rm kpc^{-2}}$. The background GC candidates have a luminosity function that closely matches the ones within 3 $R_{\rm e}$ of our dwarfs, strongly suggesting that they represent an intracluster GC population. We have verified that the background shows no significant variation across the sky regions of our dwarfs, probably because these fields are far from the Perseus cluster center.
In the case when there are masked regions within 6 $R_{\rm e}$ because of the bright contaminants, an additional correction for missing detections is applied. The uncertainty is derived from bootstrap resampling of all the GC candidates selected based on their color and concentration.

Observations of LSB dwarfs are primarily limited by the sky for ground-based telescopes, so the detection of GCs is not sensitive to the galaxy SB, until $\langle \mu_{g} \rangle_{\rm e} \lesssim 22.5$ where a significant completeness effect starts to appear for our HSC imaging. Also, brighter dwarfs show more high-SB substructures, making GC identification more difficult. Therefore, the GC measurements for those bright or high SB dwarfs are not in the scope of this work. In addition, galaxy nuclei are not expected to significantly affect our GC measurements. Not only have they been included as an extra component in the GALFIT modeling, but our inspection of non-nucleated dwarfs also shows that GCs are rarely located near the center.

\begin{figure}
\centering
\includegraphics[width=0.472\textwidth]{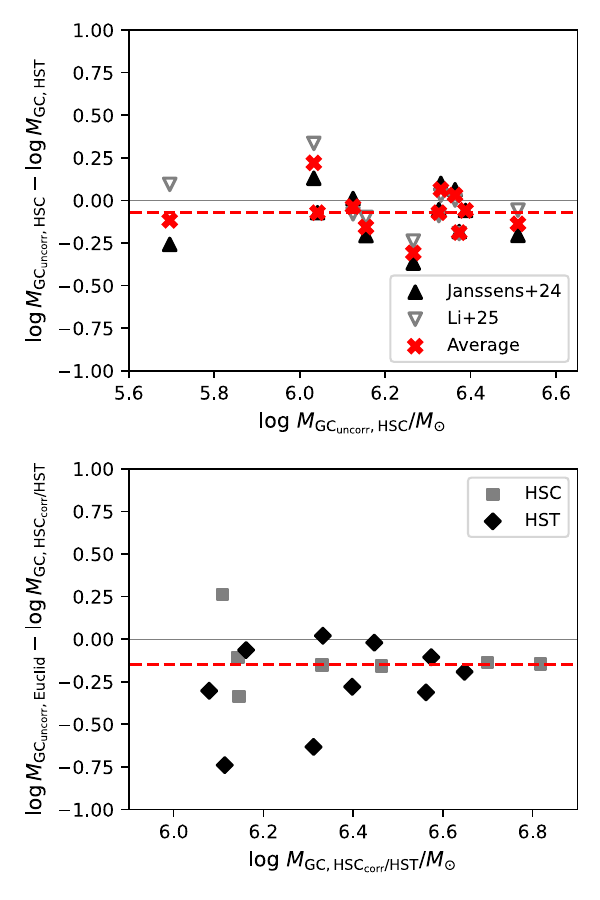}
\caption{{\it Top}: Comparison between our $M_{\rm GC}$ measurement with the HST results in the literature, where log-mass difference is plotted versus log mass. The black solid upward and gray open downward triangles show comparisons to \cite{Janssens2024} and \cite{Li2025}, respectively, while the red crosses represent the average of the two HST studies. The red horizontal dashed line shows the median difference between our HSC results and the average HST values (0.07 dex). {\it Bottom}: Comparison between the Euclid $M_{\rm GC}$ and the averaged HST values (black diamonds) or our corrected HSC values (gray squares). The Euclid values represent the average of \cite{Marleau2025} and \cite{Saifollahi2025}. The red horizontal dashed line shows the median offset of the average Euclid values (0.15 dex).}
\label{fig:mgc_comparison}
\end{figure}

To validate our $M_{\rm GC}$ estimation method and make calibrations, we compare our results for 12 relatively GC-rich LSB dwarfs with previous HST studies by \cite{Janssens2024} and \cite{Li2025}, who analyzed the same datasets using similar GC selection methods. Although these studies primarily reported the GC numbers rather than masses, $M_{\rm GC}$ can be estimated by adopting an average GC mass of $10^5\ M_{\odot}$ following \cite{Harris2017} and \cite{Janssens2024}. As presented in Figure \ref{fig:mgc_comparison} (top panel), we can see that our $M_{\rm GC}$ values derived from HSC agree well with either of the HST studies. The scatter decreases further to 0.13 dex when the HST results are averaged, yielding a median offset of 0.07 dex. We then apply this 0.07 dex upward correction to our $M_{\rm GC}$ values from HSC. %We also have good agreement with HST results in GC-poor dwarfs.

\cite{Marleau2025} and \cite{Saifollahi2025} have also performed GC measurements for Perseus dwarfs covered by Euclid. \cite{Saifollahi2025} used a traditional method similar to ours and to \citet{Janssens2024}, i.e., combining color and concentration. In contrast, \cite{Marleau2025} did not use a color constraint. Nevertheless, their results are largely consistent with each other, except that \cite{Marleau2025} appeared to find systematically lower GC numbers, which was discussed in \cite{Saifollahi2025}. We cross-match the two catalogs and select galaxies with $N_{\rm GC} \geq 10$ in both studies, finding a typical $N_{\rm GC}$ offset between them of 0.15 dex. We apply a halfway correction to the \cite{Marleau2025} and \cite{Saifollahi2025} values respectively to eliminate this offset, and then take their average. Similarly to what we have done for the HST results, we simply multiply the Euclid GC numbers by $10^5\ M_{\odot}$ to estimate $M_{\rm GC}$. Repeating the HSC to HST correction process mentioned above, we select 19 galaxies with $\log (M_{\rm GC}/M_{\odot}) \gtrsim 6$ and compare the averaged Euclid results with HST, and with HSC if no HST data are available. We find that the averaged Euclid values are 0.15 dex lower in median, and the scatter is about 0.15 dex (Figure \ref{fig:mgc_comparison}, bottom panel). Therefore, a 0.15 dex upward correction is applied to the Euclid values.

By comparing different studies, the systematic errors associated with different data and measurement methods should be the main contributors to the uncertainty in $M_{\rm GC}$. Therefore, we estimate that the typical uncertainty of our $M_{\rm GC}$ values is approximately 0.15 dex based on Figure \ref{fig:mgc_comparison}. For our final $M_{\rm GC}$ values, we preferentially use the averaged values from the two HST studies, and if HST is not available, we adopt the corrected Euclid values, and finally use our corrected HSC values where neither HST nor Euclid results are available.

Based on Eq. (13) in \cite{delosReyes2025}, we estimate the stellar masses of our dwarfs from their individual $g$-band magnitudes and $g-r$ colors. Then we obtain the GC-to-galaxy mass fraction $M_{\rm GC}/M_*$, or specific mass $S_M$ ($\equiv 100 \times M_{\rm GC}/M_*$; \citealt{Peng2008}), with values listed in Table \ref{tab:data_properties}.

\section{Results} \label{sec:results}

Having measured various properties of Perseus dwarf galaxies and their GC systems, we now analyze trends that will be used in Section~\ref{sec:discussion} to make inferences about their evolutionary histories. In the subsections below, we investigate various potential correlations between GC-richness and other galaxy properties.

\subsection{GC richness and connection to host properties}\label{sec:host}

\begin{figure}
\centering
\includegraphics[width=0.472\textwidth]{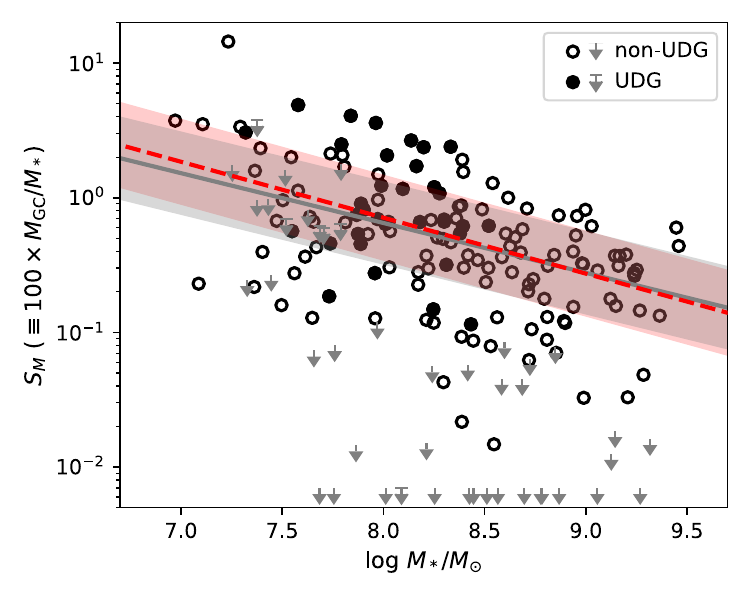}
\caption{The GC specific mass (percent mass fraction) dependence on galaxy mass of our dwarfs (open circles for non-UDGs and solid circles for UDGs). The galaxies with non-positive net GC mass within 3 $R_{\rm e}$ are shown with their 1-$\sigma$ upper limits, and the arrows at the bottom mean non-positive upper limits. The red dashed line represents the fit of $S_M$ with maximum likelihood linear regression. The red shaded band show the intrinsic scatter of the galaxies (after accounting for measurement errors). The gray solid line traces the universal stellar mass--halo mass relation from \cite{Zaritsky2023} with the scatter shown as the gray shaded band, assuming $M_{\rm GC}/M_{\rm h}=2.9 \times 10^{-5}$ \citep{Harris2017}. Our observational result is consistent with the \cite{Zaritsky2023} relation.}
\label{fig:mgc_mass}
\end{figure}

\begin{figure}
\centering
\includegraphics[width=0.472\textwidth]{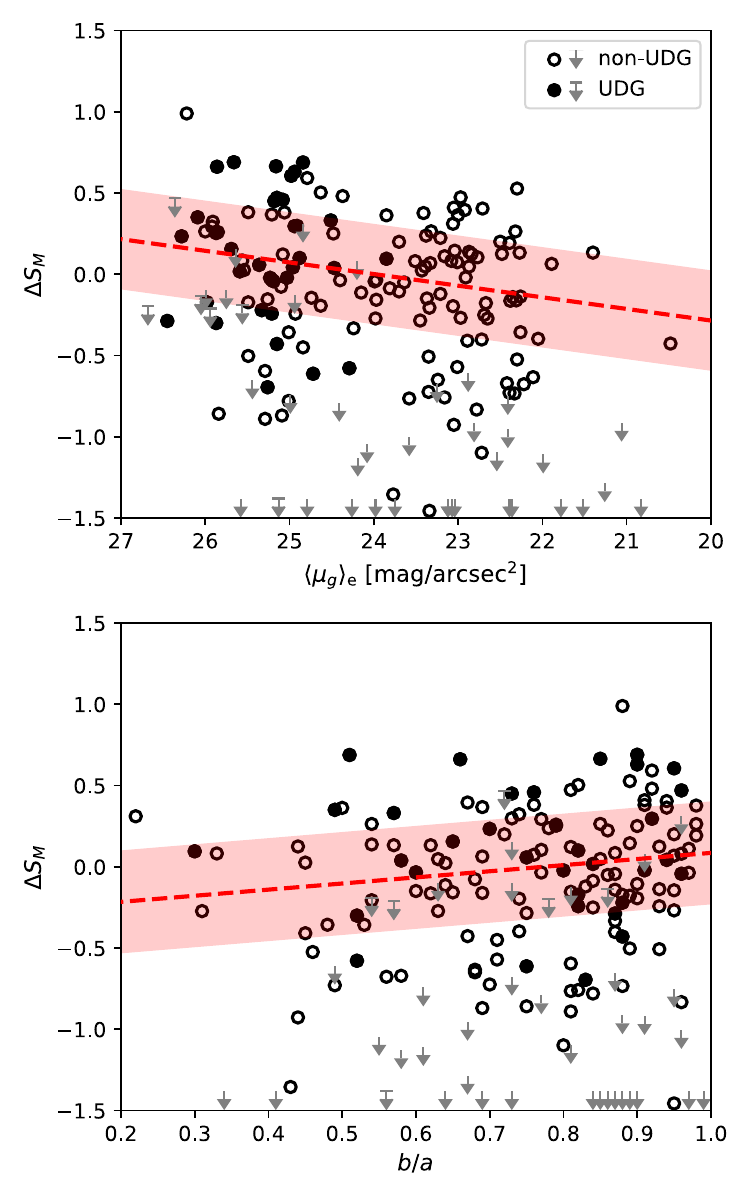}
\caption{The mass-independent GC specific mass $\Delta S_M$ dependency on $g$-band average SB (top) and on axis ratio (bottom) of the dwarfs in our sample (open circles for non-UDGs and solid circles for UDGs). The galaxies with non-positive net GC mass within 3 $R_{\rm e}$ are shown with their 1-$\sigma$ upper limits, and the arrows at the bottom mean non-positive upper limits. The red dashed lines represent the fit of $\Delta S_M$ with maximum likelihood linear regression. The red shaded bands show the intrinsic scatter of the galaxies (after accounting for measurement errors). There is a weak but significant correlation that dwarfs with lower SB are more GC-rich at fixed mass. Rounder dwarfs (higher $b/a$) seem more GC-rich, but the trend is not significant.}
\label{fig:mgc_sb_ba}
\end{figure}

\begin{figure}
\centering
\includegraphics[width=0.472\textwidth]{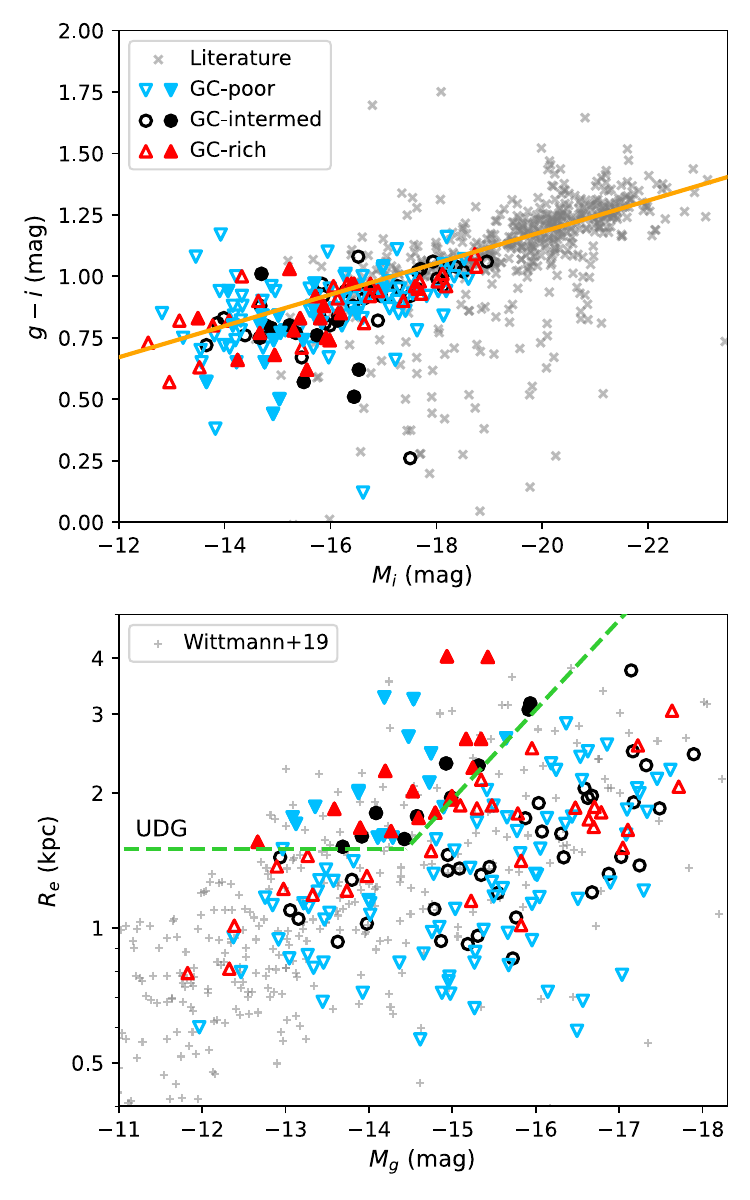}
\caption{{\it Top}: Color--magnitude distributions of Perseus dwarfs from our sample (open symbols for non-UDGs and filled symbols for UDGs; blue downward triangles, black circles, and red upward triangles represent GC-poor, intermediate, and rich) compared to other Perseus galaxies from the literature (gray crosses). The orange line is the red sequence fit to the literature galaxies, and $\Delta(g-i)$ will be used later as the residual relative to this line.
{\it Bottom}: Size--magnitude distributions of our dwarfs. The green dashed line shows the lower boundary of the UDG region. The Perseus dwarfs in \cite{Wittmann2019} are shown as gray plus signs. GC-poor dwarfs have a larger scatter in color residual $\Delta(g-i)$, but the difference is not statistically significant. GC-poor dwarfs also show a wider range of $R_{\rm e}$ at the same magnitude, and GC-rich dwarfs are larger on average.}
\label{fig:cmd_sizemag}
\end{figure}

Figure \ref{fig:mgc_mass} illustrates the correlation between GC richness and galaxy stellar mass. Despite the scatter, lower-mass dwarfs tend to exhibit higher average specific mass, consistent with known trends for the dwarf galaxies, including in other galaxy clusters \citep{Peng2008,Lim2018,Liu2019,Forbes2020}.

To quantitatively check whether the GC richness of Perseus dwarfs aligns with other dwarfs in general, we convert the stellar mass--halo mass relation for low-mass galaxies from \cite{Zaritsky2023} into an $S_M$--$M_*$ relation, assuming a constant GC-to-halo mass ratio of $M_{\rm GC}/M_{\rm h}=2.9 \times 10^{-5}$ \citep{Harris2017}. The \cite{Zaritsky2023} relation was largely based on cluster dwarfs, making it a good comparison here. Within the stellar mass range of our sample, our results are consistent with \cite{Zaritsky2023}, who also found no significant variation in this relation among four nearby galaxy clusters (Virgo, Hydra, Fornax, and Coma). The slope of our relation also agrees well with that of \cite{Saifollahi2025}, who investigated GCs in a large sample of Perseus dwarfs, although we find a different intercept. We note that discrepancies in the assumed values of $M_{\rm GC}/M_{\rm h}$, average GC mass, and mass-to-light ratio can lead to systematic offsets in the absolute values of $S_M$ (e.g., \citealt{Burkert2020}).

Next considering the UDGs in our sample, we find that they are on average more GC-rich than other dwarfs of similar stellar mass (see filled versus open circles in Figure~\ref{fig:mgc_mass}), suggesting that GC richness correlates not only with mass but also with SB. \cite{Saifollahi2025} found a similar trend in the Euclid dwarf sample in Perseus. \cite{Buzzo2025} also implicitly provided a hint of this correlation across various environments. To investigate further, we first define a relative GC richness that removes the dependence on mass. Using maximum likelihood linear regression, we derive the best-fit relation between $S_M$ and $M_*$:
$$\log S_M=(-0.42 \pm 0.02) \times \log (M_*/M_{\odot})+(3.20 \pm 0.17)$$
with an intrinsic scatter of 0.32 dex. Note that the dwarfs with non-positive $S_M$ values are incorporated into the fit as left-censored data points. We then define the relative GC richness, $\Delta S_M$, as the residual of $\log S_M$ from this relation. Furthermore, considering our typical uncertainty in $S_M$ of about 0.15 dex (see Section \ref{sec:gc_richness}), we classify 44 dwarfs with GC-richness $\Delta S_M > 0.15$~dex as GC-rich, and 99 dwarfs with $\Delta S_M < -0.15$~dex as GC-poor. The remaining 46 dwarfs can be considered GC-intermediate or ambiguous. 

For the UDGs in our sample, 15 of them are GC-rich, 15 are GC-poor, and the remaining 10 are GC-intermediate or ambiguous. In subsequent analyses and plots, we examine UDGs alongside the full dwarf sample to see if there are any indications that they are a distinct population with a different evolutionary history.

We note that previous studies usually classified dwarfs as GC-rich and GC-poor either based on absolute GC number or using a relative way (such as specific frequency or mass). However, given the wide mass range in our sample, both approaches introduce biases: the former tends to skew the GC-rich subsample toward the high-mass end, while the latter does the opposite. Therefore, we adopt a GC-richness definition that lies between the two traditional ones and is insensitive to stellar mass, as mentioned above.

We show the correlation between $\Delta S_M$ and SB in the top panel of Figure \ref{fig:mgc_sb_ba}. At fixed mass, dwarfs with lower SB tend to be more GC-rich. This correlation is not strong but is statistically significant, as indicated by a Spearman correlation coefficient of 0.20 and a $p$-value of 0.01. The slope of the best-fit linear relation is $0.07 \pm 0.01$. \cite{Saifollahi2025} reached a similar conclusion from the Euclid sample, based on a two-sample comparison between high- and low-SB dwarfs. Here we take one step further by showing the continuous correlation between SB and GC richness. We also achieve similar results if applying our method to the Euclid sample (see Appendix \ref{sec:euclid}). As discussed in Section \ref{sec:gc_richness}, we are still able to reliably detect GCs in the high-SB galaxies in our sample. Therefore, the trend between SB and $\Delta S_M$ here should not be driven by completeness effects.

\begin{figure*}
\centering
\includegraphics[width=0.7\textwidth]{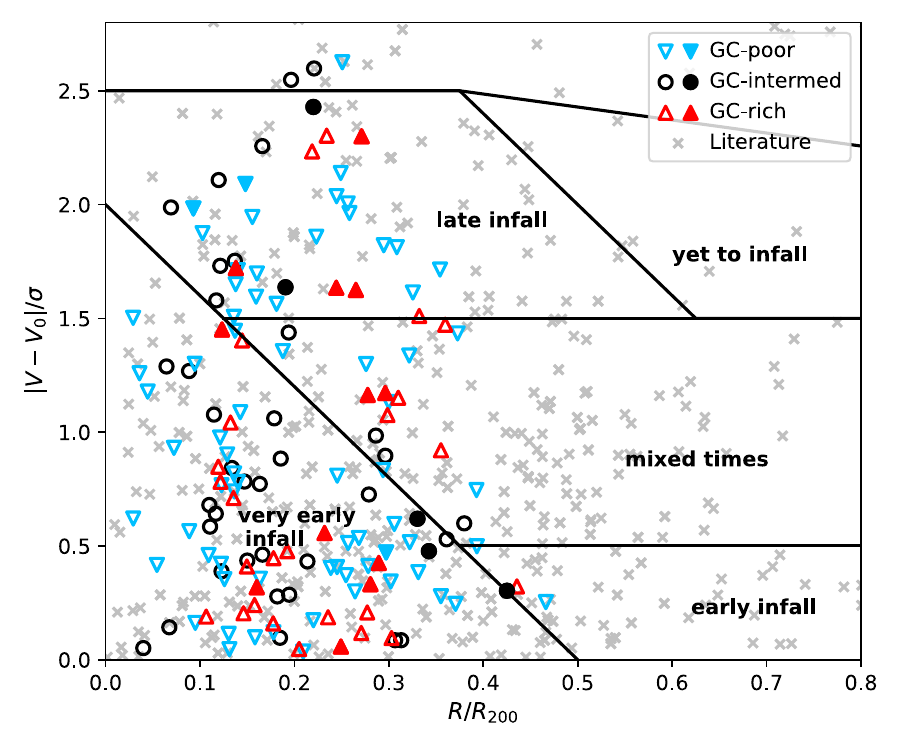}
\caption{Position--velocity phase-space plot of Perseus galaxies. Our sample of dwarfs is plotted with open symbols for non-UDGs and filled symbols for UDGs; blue downward triangles, black circles, and red upward triangles represent GC-poor, intermediate, and rich. The gray crosses are other Perseus galaxies in the literature. The horizontal axis represents the projected clustercentric radius normalized by the virial radius $R_{200}$, and the vertical axis shows the recessional velocity relative to the cluster center $|V-V_0|$ and normalized by the cluster velocity dispersion $\sigma$. The typical uncertainty on the $y$-axis is $\sim$ 0.01, and for very faint dwarfs can be as large as $\sim$ 0.05. The infall-time regions from \cite{Rhee2017} are overplotted. GC-poor dwarfs are dominant in the later infall regions, while GC-rich dwarfs are slightly more common than GC-poor in the earlier infall regions.}
\label{fig:phase_space}
\end{figure*}

When directly comparing $S_M$ and SB, the correlation is stronger, with a correlation coefficient of 0.58. In comparison, the correlation coefficient between mass and $S_M$ is $-0.47$, and the scatter is almost the same. \cite{Lim2018} and \cite{Forbes2020} found a similar trend for Coma cluster dwarfs including UDGs, although their high-SB galaxies are much more concentrated toward the cluster center than the rest of the sample.

We next compare dwarf galaxy photometric axis ratios ($b/a$) with $\Delta S_M$ in the bottom panel of Figure \ref{fig:mgc_sb_ba}. The rounder dwarfs appear to be more GC-rich than the more elongated ones, but this trend is weak and not highly significant, with a Spearman correlation coefficient of 0.15 and a $p$-value of 0.08. For UDGs, we do not see a significant correlation ($p = 0.8$), which is in line with the finding in \cite{Pfeffer2024} for a smaller sample of Perseus UDGs. For the Euclid sample, the significance of the correlation between $b/a$ and relative GC richness becomes even weaker (see Appendix \ref{sec:euclid}). These results for Perseus contrast with findings in other environments, where GC-rich UDGs were found to be significantly rounder than GC-poor ones \citep{Lim2018,Pfeffer2024,Buzzo2025}. 

To begin understanding the stellar populations of our dwarfs, we present their color--magnitude distributions in Figure \ref{fig:cmd_sizemag} (upper panel), along with a comparison sample of confirmed Perseus members in the literature (\citealt{Kang2024}; based on SDSS colors, using the same calibration as our HSC imaging). Our dwarfs generally follow the same red sequence as the literature ones (green dashed line), with a small offset toward bluer colors (which is probably an artifact of imperfect sky subtraction; see Section \ref{sec:galfit}). We define a color residual $\Delta(g-i)$ by subtracting the red sequence trend, which is fitted with the confirmed cluster members in \cite{Kang2024}:
$$g-i=-0.064\times M_i-0.094$$
We do not see a statistical difference in $\Delta(g-i)$ between the GC-rich and -poor dwarfs, based on a Kolmogorov--Smirnov test with a $p$-value of 0.8. Note that variations in stellar populations can be difficult to detect with a single optical color, owing to age--metallicity degeneracies.

Figure \ref{fig:cmd_sizemag} (bottom panel) also shows the size--magnitude distribution of our dwarfs. At a given luminosity, the GC-rich dwarfs tend to have larger sizes, reflecting their more diffuse morphology and lower SB. In contrast, the GC-poor dwarfs exhibit a wider range of sizes at the same magnitude. A similar pattern is found for the Euclid sample in Figure \ref{fig:euclid_gc}.

\subsection{GC richness and connection to infall stages}\label{sec:infall}

\begin{figure}
\centering
\includegraphics[width=0.472\textwidth]{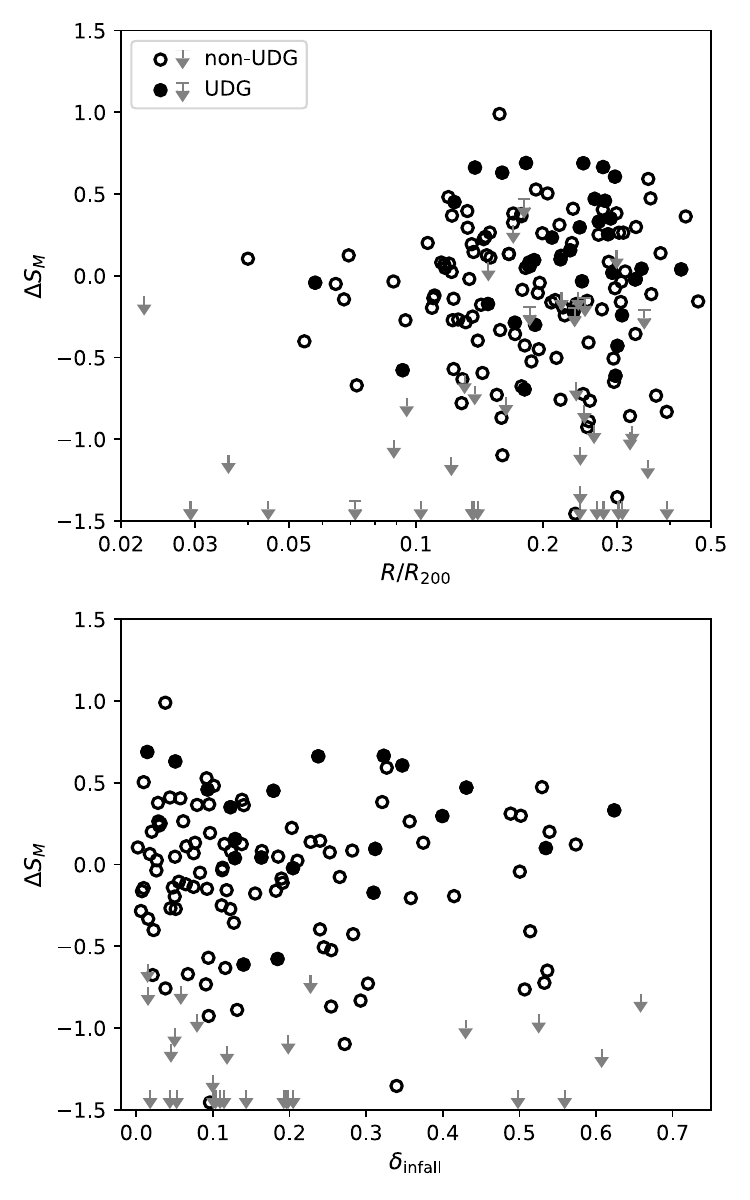}
\caption{The mass-independent GC specific frequency $\Delta S_M$ dependence on projected clustercentric radius (top) and on infall-time parameter (bottom, $\delta_{\rm infall}=R/R_{200} \times |V-V_0|/\sigma$) of the dwarfs in our sample (symbols as in Figure~\ref{fig:mgc_sb_ba}). The bottom panel has fewer points than the top because it only includes the dwarfs with radial velocity measurements. No significant monotonic correlation is observed between $\Delta S_M$ and $R/R_{200}$ or $\delta_{\rm infall}$.
}
\label{fig:mgc_infall}
\end{figure}

\begin{figure}
\centering
\includegraphics[width=0.472\textwidth]{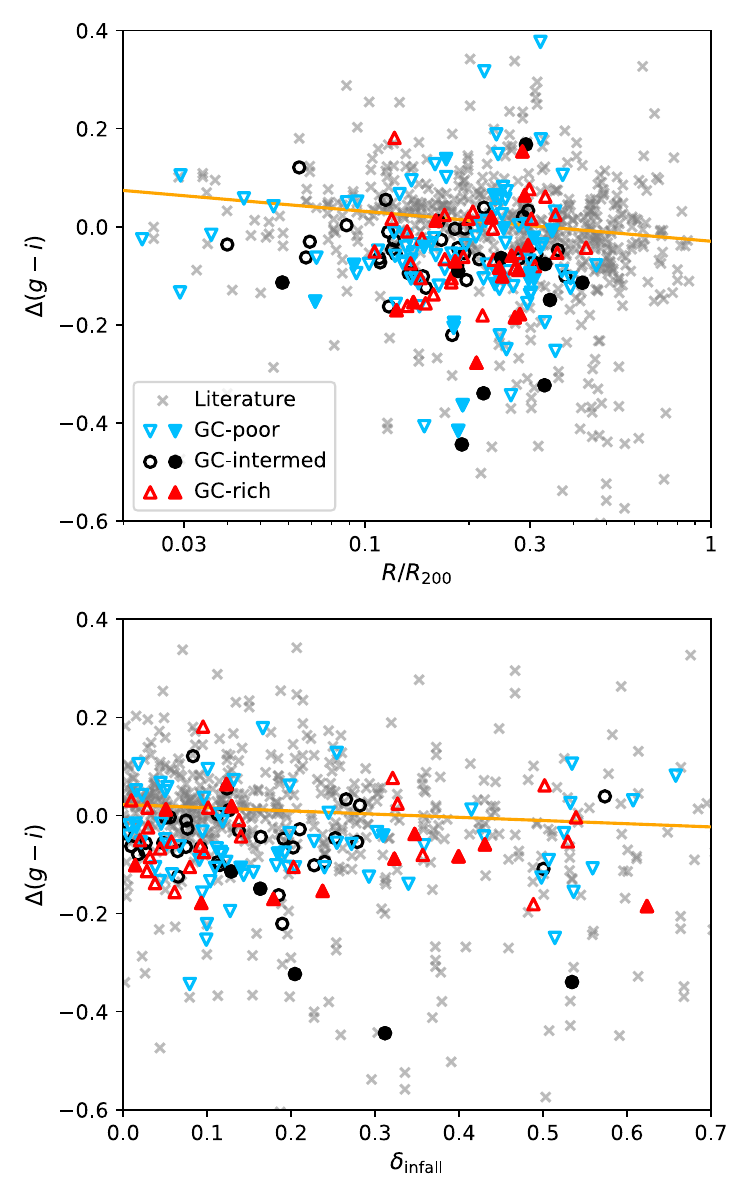}
\caption{The $\Delta (g-i)$ galaxy residual color dependence on projected clustercentric radius (top) and on $\delta_{\rm infall}$ (bottom) for Perseus galaxies. 
Our sample of dwarfs is plotted with open symbols for non-UDGs and filled symbols for UDGs; blue downward triangles, black circles, and red upward triangles represent GC-poor, intermediate, and rich.
The gray crosses are other Perseus galaxies in the literature, and the orange solid lines show the linear fits to these data points. The bottom panel has fewer points than the top because it only includes the dwarfs with radial velocity measurements. Our sample dwarfs are generally consistent with the literature galaxies, which become bluer at larger clustercentric radius and later infall time, while showing no significant difference between GC-rich and GC-poor galaxies. 
}
\label{fig:color_infall}
\end{figure}

We show the distribution of our dwarf galaxies in phase-space in Figure \ref{fig:phase_space}, i.e., the recessional velocity relative to the cluster normalized by the cluster velocity dispersion, versus the projected clustercentric radius normalized by the cluster virial radius. We overlay the boundaries of the characteristic regions of projected phase-space identified in simulations of galaxy clusters \citep{Rhee2017}, labeling each region by the dominant infall group. Here we note that projection effects lead to a wide range of times since infall for each region, with mean times of $\sim$~6~Gyr for the ``very early'' region, and $\sim$3--4~Gyr for the early/mixed/late regions.
Most of our galaxies are located within the very early infall and late infall regions, and a few are in the mixed times region.

No significant difference in velocity distribution is observed between UDGs and non-UDGs, with a $p$-value of 0.3, suggesting they may have similar infall properties given the limited radial coverage of our sample ($R/R_{200} \lesssim 0.5$). This result is similar to the Coma cluster, which has no clear evidence for UDGs being kinematically distinct from other dwarfs \citep{Alabi2018}. However, in Hydra I, UDGs appear to have lower relative velocities on average \citep{Forbes2023}.%, which might be due to a selection effect.

A simple two-region split of the phase space at $|V - V_0|/\sigma = 1.25$ indicates that the relative fraction of GC-poor dwarfs increases toward the later-infall region, where they make up about 55\% of the population, compared to only 20\% for GC-rich ones. In the earlier infall region, GC-rich dwarfs become relatively more common, accounting for 28\% of the population versus 44\% for GC-poor ones. Although these contrasts are not statistically significant, the overall trend is consistent with expectations.

Among the three infall regions that our dwarfs primarily occupy, we find that dwarfs in the mixed times region tend to be more GC-rich, with a median $\Delta S_M=-0.00 \pm 0.26$. In contrast, those in the late infall region are typically more GC-poor, with a median $\Delta S_M=-0.19 \pm 0.22$. The very early infall dwarfs fall in between, with a median $\Delta S_M=-0.14 \pm 0.05$. We find a similar non-monotonic trend for UDGs. However, bootstrap-based uncertainty estimates indicate that the differences in median $\Delta S_M$ across the three infall regions are not statistically significant. Expanding the sample size would allow for a better assessment of this trend, especially in the relatively late infall regions.

For a more continuous comparison between GC richness and infall time, Figure \ref{fig:mgc_infall} (top panel) shows the correlation between $\Delta S_M$ and projected clustercentric radius. We do not see any significant trend, as indicated by a Spearman correlation $p$-value of 0.8. A similar test for the Euclid sample gives a $p$-value of 0.5 (see Appendix \ref{sec:euclid}), reinforcing the indication from \cite{Marleau2025} that the GC-rich Perseus dwarfs are not preferentially located near the cluster center. These Perseus results differ from observations of Virgo \citep{Peng2008} and Coma \citep{Lim2018}, where GC richness was found to depend significantly on clustercentric radius. However, the dwarf samples in those studies had much larger radial coverage, extending to approximately one virial radius. 

When incorporating velocity into consideration, we define $\delta_{\rm infall}$ as the product of the $x$- and $y$-axis values in Figure \ref{fig:phase_space} (i.e., $\delta_{\rm infall} \equiv R/R_{200} \times |V-V_0|/\sigma$; see \citealt{Alabi2018,Forbes2023}), which should be a better indicator of infall stage than using radius alone. However, no significant correlation between $\Delta S_M$ and $\delta_{\rm infall}$ is found, with a Spearman correlation test $p$-value of 0.7. Using a similar analysis, the UDGs do not exhibit significant correlations of GC richness with either clustercentric radius or $\delta_{\rm infall}$. We note that the Spearman test is designed for monotonic trends, so the possible non-monotonic pattern with infall stage mentioned above may not be reflected in this statistic.

In Figure \ref{fig:color_infall}, we present the correlations between galaxy residual color $\Delta (g-i)$ and projected clustercentric radius and $\delta_{\rm infall}$. Overall, the Perseus galaxies exhibit a clear trend of bluer $\Delta (g-i)$ colors with increasing clustercentric radius and $\delta_{\rm infall}$, consistent with younger stellar populations at later infall stages. The dwarfs in our sample are generally consistent with this trend, although with shallower slopes. No significant difference can be seen between the GC-rich and GC-poor dwarfs. A similar trend was found for Coma dwarfs and UDGs by \citet{Alabi2018}, although with a slope that appears to be several times steeper than in Perseus (after using the conversion between $g-i$ and $B-R$ colors in \citealt{Usher2012}). Other clusters in the literature have shown less consistent radial color trends for UDGs (see discussion in \citealt{Thuruthipilly2024}).

\begin{figure}
\centering
\includegraphics[width=0.472\textwidth]{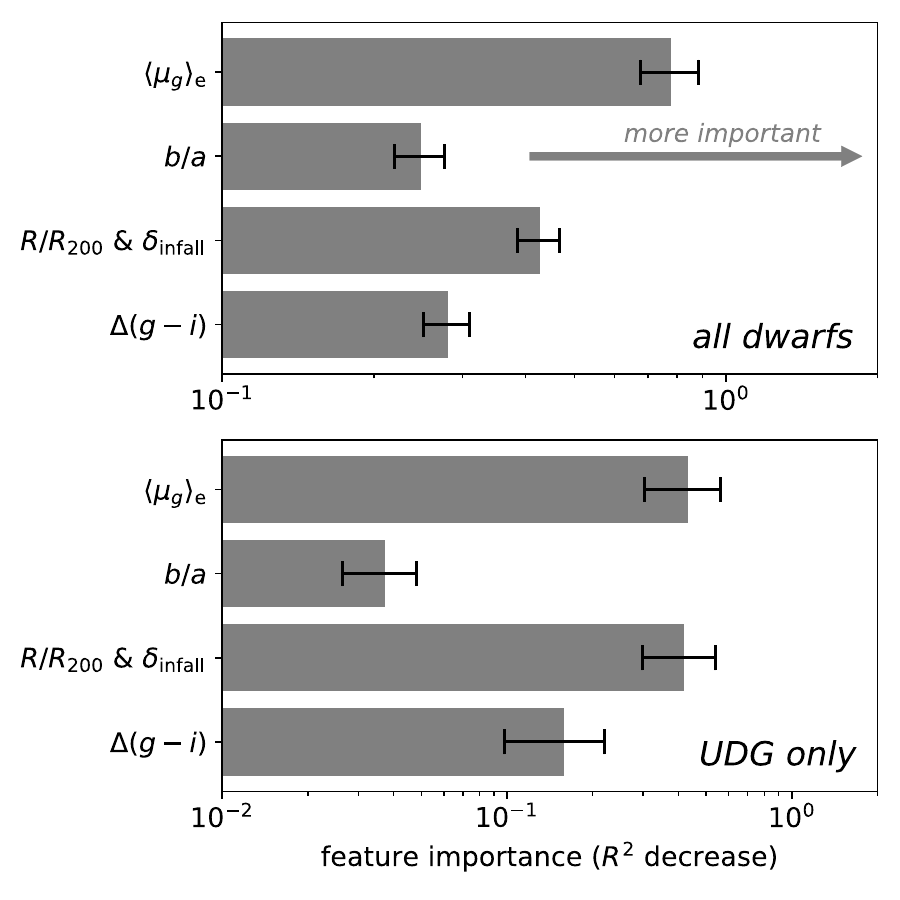}
\caption{Permutation feature importance for modeling $\Delta S_M$ based on various galaxy properties, shown for our dwarf galaxies with available velocity information (top) and for the UDGs among them (bottom). A random forest is applied here to model $\Delta S_M$, and the feature importance is quantified by the decrease in model score $R^2$ caused by randomly permuting the corresponding feature. $R/R_{200}$ and $\delta_{\rm infall}$ are grouped during permutation due to the relatively strong correlations between them. Average SB shows the highest importance, followed by infall stage, color, and axis ratio. The pattern is similar for UDGs, although the importance of SB decreases due to their relatively narrower SB range.
}
\label{fig:feature_importance}
\end{figure}

\subsection{Which properties correlate better with GC richness?}

Here we conduct an assessment based on permutation importance \citep{Breiman2001} to better quantify the impact of each property on $\Delta S_M$, as shown in Figure \ref{fig:feature_importance}. First, we model $\Delta S_M$ with a random forest. We then randomly shuffle the values of each feature, considering a feature more important if this permutation reduces the model’s $R^2$ score more, where $R^2$ measures the proportion of variance in the target variable explained by the model. However, considering the relatively strong correlations between $R/R_{200}$ and $\delta_{\rm infall}$, we group some of them together while doing the random permutation. Average SB shows the highest importance, followed by infall stage, color, and axis ratio. For UDGs, the pattern is similar, although the importance of SB decreases due to their relatively narrower SB range.

At first glance, these results may appear inconsistent with those in Section \ref{sec:results}, where infall stage and color (or stellar population) do not show significant correlations with GC richness. This suggests a limitation of the Spearman correlation test, which is sensitive only to monotonic correlations. In contrast, infall stage and color may have more complex and non-monotonic correlations with $\Delta S_M$ that can be better captured by the random forest model. As discussed in Section \ref{sec:infall}, from very early infall to late infall, the median $\Delta S_M$ shows a trend of first increasing and then decreasing. Also, we see that GC-poor dwarfs exhibit a larger color scatter in Figures \ref{fig:cmd_sizemag} and \ref{fig:color_infall}. These trends are not simply monotonic, although neither is statistically significant, which is likely due to the limited sample size or simplified trends examined.

We also attempted to identify subpopulations of our dwarfs in the high-dimensional space defined by all the properties discussed above, e.g., to look for a distinct population of UDGs, but were unable to obtain stable or meaningful results. This may suggest that more information about stellar population, including age, metallicity, and even star formation history, is crucial if we need to further distinguish between those potential subpopulations \citep{Buzzo2025}. For example, subpopulations of GC-rich UDGs may be distinguished primarily by their metallicities \citep{FerreMateu2023,FerreMateu2025,Levitskiy2025}. It would also be helpful if dwarfs could be included in the analysis at even lower SB (which are challenging to detect and characterize; e.g., \citealt{Li2025b}).

\section{Discussion} \label{sec:discussion}

\begin{figure*}
\centering
\includegraphics[width=1\textwidth]{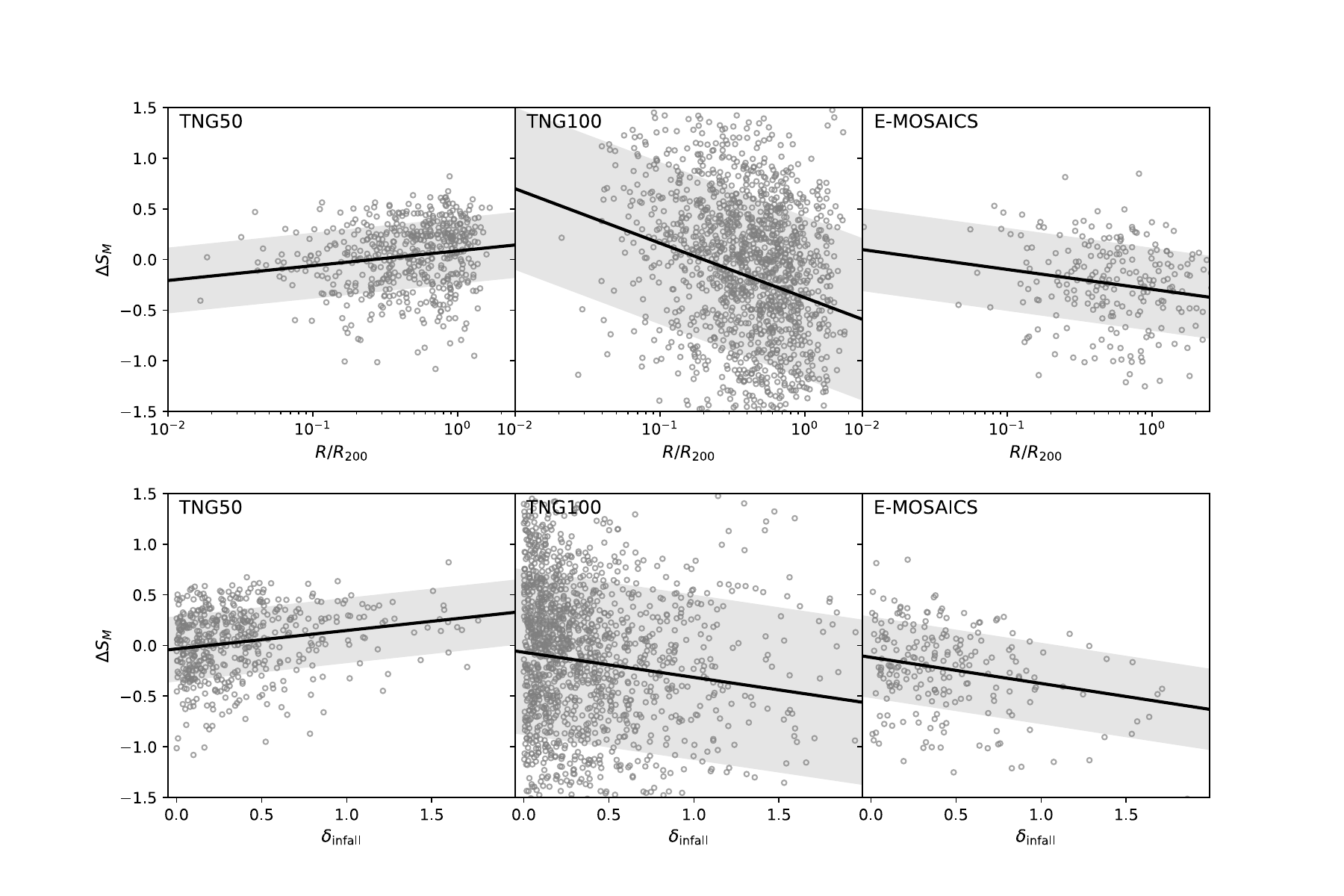}
\caption{The mass-independent GC specific frequency $\Delta S_M$ dependence on galaxy clustercentric radius (top row) and $\delta_{\rm infall}$ (bottom row) for TNG50, TNG100 and E-MOSAICS simulations. The black lines represent the fit of $\Delta S_M$ with maximum likelihood linear regression for the simulation data points. While all three simulations predict significant trends of $\Delta S_M$ with clustercentric radius and $\delta_{\rm infall}$, TNG50 shows the opposite correlations compared to TNG100 and E-MOSAICS.
}
\label{fig:simulation_infall}
\end{figure*}

\subsection{Insight from simulations}

In the previous section, we have shown how the mass-independent GC-richness metric $\Delta S_M$ correlates with various properties for our sample of Perseus dwarf galaxies. Before diving into the discussion of a broader picture of Perseus dwarfs and GC evolution, it is helpful to consider what is predicted by current theory. Understanding the formation and evolution of GCs within galaxies in a full cosmological context remains a fundamental challenge for theory. Numerical simulations of the full range of physical scales involved, particularly in the context of high-mass galaxy clusters, are not yet feasible. Instead, prescriptions for subgrid physics are adopted within the simulations, based on what is known about star cluster physics in the Local Universe. The models vary in their prescriptions, and comparisons of their predictions to observations can further inform the models and our understanding of GC systems.

In this discussion, we will focus on correlations between GC richness and infall stage, using the results from three major simulations that incorporate GCs within dwarf galaxies within galaxy clusters: Illustris TNG50 \citep{Pillepich2019,Nelson2019,Doppel2023}, 
TNG100 \citep{Carleton2021},
and E-MOSAICS \citep{Pfeffer2018,Kruijssen2019,Pfeffer2024}. The latter two simulations do not adequately resolve the internal properties of dwarf galaxies, and therefore we do not consider correlations with size, SB, and axis ratio. All of these simulations incorporate two basic effects: dwarfs that fall in to the cluster earlier have higher $S_M$, followed by a decrease in $S_M$ with time.

TNG50 provides a combination of the highest-resolution simulation, and the most simplified treatment of GCs \citep{Doppel2023}. We choose the most massive galaxy cluster in the simulation ($M_{200} \sim 2\times 10^{14} M_{\odot}$), which is closer to the Virgo cluster mass than to Perseus. We select dwarf galaxies with stellar masses between $10^7$ and $10^9\ M_{\odot}$ to match the mass range of our observed sample. In this simulation, GCs with fixed mass were assigned to galaxies using a tagging scheme based on an empirical relation 
between GC system mass and host halo mass. Tidal stripping has the main subsequent impact on the GC system mass of individual galaxies.

The GC modeling in TNG100 focuses on dwarfs within three galaxy clusters of mass $M_{200} \sim 3\times 10^{14} M_{\odot}$ \citep{Carleton2021}. These TNG100 dwarfs have a mass range of $7<\log M_*/M_{\odot}<9$. In this simulation, GC formation is closely tied to intense star formation, with specified mass function and spatial distribution based on observation. In addition to tidal stripping, tidal disruption and evaporation within the galaxy are also accounted for at all times.

From E-MOSAICS \citep{Pfeffer2024}, we include dwarfs with stellar mass of $7.5<\log M_*/M_{\odot}<9$, selected in four massive  halos with $M_{200}>3 \times 10^{13} M_{\odot}$. Compared to the other two simulations and the Perseus cluster, these halos are much less massive, and more similar to galaxy groups than clusters. Nevertheless, they still provide valuable insights due to their more comprehensive modeling. In particular, E-MOSAICS treats star clusters as subgrid components of baryonic particles in their cosmological simulations, leading the GC formation and evolution to be more tightly linked to local properties than in TNG100.

We apply the same formula used in our Perseus study to compute $\Delta S_M$ for the simulated dwarfs, and plot the correlations with $R/R_{200}$ and with $\delta_{\rm infall}$ in Figure \ref{fig:simulation_infall}. These are the quantities that can be directly compared to observations, although we have also checked the corresponding three-dimensional quantities in the simulations and found the same trends. Furthermore, for TNG50 we have direct infall time measurements available, and also find that these track the infall-time proxy parameters above.

We see that in both TNG100 and E-MOSAICS, GC richness decreases significantly on average from early to late infall (based on Spearman correlation tests) -- as previously shown in Figure~1 of \cite{Carleton2021} and in Figure~2 of \cite{Pfeffer2024}. The underlying physical reason is that early-infall dwarfs also started forming relatively early, when star formation rate densities were higher and thus more of the stars formed within clusters. The subsequent depletion of GCs after infall increases the scatter in $S_M$ but does not remove the main trend.

In contrast, TNG50 shows the opposite correlation, where later infalling dwarfs are typically more GC-rich.
This result is a surprise, since an earlier generation of these models suggested that GC richness would be higher at earlier infall times and at smaller clustercentric radii \citep{Ramos-Almendares2020}. It seems the predictions from this type of model are sensitive to the details of the stellar-to-halo mass relation and of the tidal stripping treatment.
Models such as TNG100 and E-MOSAICS that include additional bursts of star formation around infall may lead to more stable predictions for $S_M$ versus infall time.

To make better comparisons with our observations, we constrain the simulated data to match our sample size and infall parameter range ($R/R_{200}<0.5$ and $\delta_{\rm infall}<0.7$), and add an observational uncertainty of 0.15 dex to $\Delta S_M$. After applying these, we find that the correlation with $\delta_{\rm infall}$ in TNG100, as well as both correlations in E-MOSAICS, are no longer statistically significant. Both correlations in TNG50, and the clustercentric radius correlation in TNG100 remain significant, but this is largely driven by data points very close to the cluster center. When we further constrain the radius range to $0.1 < R/R_{200} < 0.5$, the $p$-values of these three correlations drop to the range of 0.05–0.10. While no longer highly significant, they are still noticeably more significant than those found from our observation (see Section \ref{sec:infall} and Figure~\ref{fig:mgc_infall}).

Similar to Section \ref{sec:infall}, we can also investigate the median $\Delta S_M$ trends within each infall region in Figure \ref{fig:phase_space} for the simulations. Due to the larger radius coverage, there are enough objects in the early infall region for simulations, in addition to the three regions we have discussed above for our Perseus dwarfs (i.e., very early infall, mixed times, and late infall). TNG50 shows the trend most consistent with our observations: $\Delta S_M$ increases from very early infall to mixed times and then decreases toward late infall. The median $\Delta S_M$ in the four regions is $-0.33$, $-0.06$, 0.04, and $-0.09$, respectively. Different from our observations, the median $\Delta S_M$ in the very early infall region is lower than for late infall. TNG100 once again exhibits a trend that is perfectly opposite to TNG50, with the median $\Delta S_M$ in the four regions of $-0.61$, $-1.58$, $-2.79$, and $-1.20$. We note that these values are more negative because TNG100 contains a much higher fraction of galaxies without GCs (likely due to low resolution at very early times, as discussed in \citealt{Carleton2021}). If we exclude these cases, we find a similar trend, albeit much flatter in the latter three regions. Differently from TNG50 and TNG100, E-MOSAICS shows a more monotonic trend, with $\Delta S_M$ gradually decreasing from early to late infall, and the four medians are $-0.15$, $-0.21$, $-0.21$, and $-0.42$.

It appears that none of the three simulations reproduce the observed trends of GC richness with infall time, with the caveat that the simulations are all of lower mass clusters than Perseus. Hence it is worth revisiting observational work on other clusters. \cite{Peng2008} and \cite{Lim2018} found a negative correlation between GC specific frequency and clustercentric radius for Virgo dwarfs and for Coma UDGs, respectively. These results are in qualitative agreement with the TNG100 and E-MOSAICS models, and support a conventional picture of GC richness driven by earlier infall time.

In summary, the available simulations predict that $\Delta S_M$ should exhibit a weak but statistically significant correlation with infall stage. The different models vary in their predictions, suggesting that observations of GC richness versus infall stage could in principle help to improve the models -- in particular focusing on the treatment of tidal stripping of GCs. On the other hand, the observations do not all find the same trends. For Perseus, it could be that the radial coverage is insufficient to detect a significant trend, which suggests that future work should aim to build a large sample of dwarfs with GC measurements beyond 0.5 $R_{200}$.

\subsection{Evolutionary scenarios of Perseus dwarfs and GCs}

With the multi-parameter analysis from observations and with 
guidance from simulations above, we now discuss evolutionary scenarios for Perseus dwarfs and their GC systems. From our observed sample, the strongest and most robust trend is the negative correlation between GC richness and SB. This trend is counter to a basic expectation that GCs arise in the highest densities of star formation \citep{Elmegreen1997,Goddard2010,Kruijssen2014,Ma2020} and suggests two logical alternatives. The first is that the lowest SB dwarfs today had the highest densities at the time of formation, and subsequently expanded either due to tidal heating or to internal feedback (e.g., \citealt{Danieli2022}). The second is that there is a strong correlation between galaxy density and GC disruption, i.e., all the dwarfs started out with similar $S_M$ but the highest SB ones lost more of their GCs to disruption \citep{Moreno-Hilario2024}. Here we note that \citet{Forbes2025} attempted to distinguish between these two explanations for GC variations in UDGs using stellar populations, and found indications of GC formation being the dominant effect over GC disruption (see also \citealt{Liu2016}). Obtaining more stellar population measurements of Perseus dwarfs could therefore be crucial for further progress on understanding the SB trends. Studies of stellar population gradients and angular momentum can also look for signatures of galaxy expansion (e.g., \citealt{FerreMateu2025,Levitskiy2025}).

Our permutation feature importance analysis has identified the infall stage as the second most important property correlated with GC richness. However, compared to other galaxy clusters in the nearby Universe and to predictions from simulations, the trend in the Perseus cluster appears unexpectedly weak, although this result requires further confirmation with larger samples and broader coverage in clustercentric radius for Perseus. Also, more work is needed in the similar mass Coma cluster, on a wider variety of dwarfs and not only the UDGs.

There is evidence suggesting that Perseus has experienced a major merger (e.g., \citealt{Marleau2025,HyeongHan2025}), which might have mixed up the orbits of the dwarfs, making their locations in phase space no longer reliable reflections of their true infall times. However, the Perseus merger time estimates of 6--8~Gyr are long enough ago to allow for the subsequent re-establishment of the standard infall regions of phase space. Furthermore, Virgo and Coma are also thought to be in a state of merging (e.g., \citealt{Adami2009,Boselli2014}) yet still show anti-correlations between GC richness and infall time, and Coma shows a steeper relation of color with infall time than in Perseus.
Thus the lack of a clear correlation between GC richness and infall time in Perseus remains puzzling.

We also confirm the finding of \citet{Pfeffer2024} that Perseus differs from other environments (including Coma and Virgo; \citealt{Lim2018}) in showing no significant correlation between GC richness and axis ratio. \citet{Pfeffer2024} predicted from simulations that GC-rich dwarfs would have higher gas pressures and be more dispersion-supported at formation, leading to rounder shapes. Furthermore, one might expect that early infall dwarfs also become rounder through tidal effects \citep{Barber2015}.

Thus Perseus shows {\it two} unexpected non-trends of the dwarf GC properties, with axis ratio and with infall time, which could be related. We speculate that Perseus dwarfs may have experienced more complex and diverse star formation histories and/or environmental disturbance histories compared to other galaxy clusters, which could dilute the anticipated correlations. For example, just as UDGs occupying the same parameter space of size and SB are now thought to represent a mixed bag of different origins, so also there might be distinct subpopulations of GC-rich dwarfs whose superposition confounds simple correlations. More specifically, there could be dwarfs whose GC-richness is driven by a conventional picture of early infall, as well as failed galaxies with late infall times and flattened, prolate-triaxial shapes \citep{vanDokkum2019}. Disentangling these superpositions could be done through studying additional parameters of the galaxies such as stellar populations and dynamical mass (e.g., \citealt{Beasley2016,Pfeffer2024,Buzzo2025,Levitskiy2025}),
wherein early infallers are metal-rich and depleted in dark matter,
while failed galaxies are metal-poor and have overmassive dark matter halos.
It would also be useful to check the differences in GC properties among individual galaxy clusters in cosmological simulations and their connections to merger histories, but that is beyond the scope of this work. 

It is not clear whether the larger color scatter observed in GC-poor dwarfs reflects a wider range in their age or metallicity, or both. While one may expect GC-poor dwarfs to be mainly late infallers with younger ages and bluer colors, our results do not provide strong support for this scenario. Additionally, a few dwarfs lie significantly above the red sequence, which may suggest mass loss due to tidal stripping, resulting in metallicities that are higher than expected for their surviving mass.

In addition to these overall trends, we have also identified several interesting cases. Figure \ref{fig:phase_space} shows that three GC-rich dwarfs (H555, J513793, EDwC-0986) are located in the late infall region, with $|V-V_0|/\sigma > 2.2$. We note that H555 is a UDG, and also that the previously studied late-infall UDG R24 \citep{Gannon2022} is classified as GC-intermediate. However, one should be mindful that about 30\% of the galaxies in this projected region of phase space are actually ancient infallers \citep{Rhee2017}. H555 and EDwC-0986 do exhibit bluer colors than the red sequence, consistent with expectations for late-infall galaxies. Their GC systems may have formed in relatively recent starbursts, potentially due to ram-pressure compression at infall (e.g., \citealt{Mahajan2010}) or to mergers in groups prior to infall (e.g., \citealt{Lahen2019}).

In contrast, J513793 has significantly redder colors, suggesting it might already be quenched before their infall -- in apparent tension with the standard picture where massive dwarfs are always star-forming in the field \citep{Geha2012}. Similar cases have been reported in Virgo and suggested to be caused by ``pre-processing'' in pre-infall groups \citep{Bidaran2022,Bidaran2023}, which for the Perseus dwarfs may imply early infall within those groups. Another possible explanation is the backsplash mechanism (e.g., \citealt{Balogh2000,Benavides2021}) in which dwarfs interact with a massive halo at an earlier epoch, lose their gas, and are then ejected back into the field. Additionally, ram pressure from diffuse gas in cosmic web filaments and sheets can also quench isolated dwarf galaxies \citep{Benitez-Llambay2013,Pasha2023,Benavides2025,Bidaran2025}.

\section{Conclusion} \label{sec:conclusion}

In this work, we have presented a study of 189 dwarf galaxies (including 40 UDGs) in the Perseus cluster and their GC systems. Compared to previous studies, our sample represents the largest collection of dwarf galaxies within a single galaxy cluster that includes both imaging and spectroscopic data, with the dwarfs broadly and uniformly distributed across parameter space (e.g., stellar mass, size, GC content). All our dwarfs are measured and analyzed with attention to homogeneity, which minimizes systematics. Our photometric analysis is primarily based on Subaru HSC imaging for both galaxies and their GC systems, supplemented by results from HST and Euclid available in the literature. In addition, velocity measurements from Keck spectroscopy and from previous studies enable us to investigate the infall stages of these dwarfs through phase-space analysis. Our main findings are summarized as follows:

\begin{itemize}

\item Similar to other nearby clusters, Perseus dwarfs show a significant negative correlation between GC specific mass and galaxy stellar mass, accompanied by moderate scatter. At fixed stellar mass, their GC richness is also broadly consistent with that of dwarfs in other clusters.

\item At fixed stellar mass, Perseus dwarfs with lower surface brightness or larger sizes tend to be more GC-rich. This result has been suggested in the past and demonstrated continuously here. Further work is needed to determine if the effect is caused by augmented GC formation or by inefficient GC disruption in the lower density dwarfs.

\item In Perseus, the correlation between GC richness and axis ratio of dwarf galaxies appears weaker than in other environments, where it has been interpreted as more efficient GC formation in pressure-supported dwarfs. This observation may imply more complex evolutionary histories in Perseus.

\item Different from observations in Virgo and Coma, as well as from cosmological simulations, we do not find significant correlations between GC richness and clustercentric radius or infall stage for Perseus dwarfs.
Combined with other findings, we suggest that Perseus dwarfs may represent a population with more complex and diverse evolutionary histories.
However, more complete observations in both Perseus and in other galaxy clusters are needed to further test for cluster-to-cluster variations in galaxy and GC evolutionary histories.

\item With the limited number of UDGs in our sample, we do not find significant differences between UDGs and other dwarf galaxies in Perseus across various correlations, including those between GC richness and axis ratio, infall indicators, or color. UDGs are generally more GC-rich, which may be a reflection of a continuous trend with SB. Further tests of internal dynamics and stellar populations may be needed to reveal distinctions between UDGs and other dwarfs.

\end{itemize}

We emphasize that the similarities and differences in dwarf galaxies and their GC systems among nearby galaxy clusters provide important benchmarks for theories. For the Perseus cluster, future work should focus on building similar samples beyond 0.5 $R_{200}$. Expanding the spatial coverage will enable more robust testing of the trends studied in this work and provide deeper insight into the nature of dwarf galaxies and their GC systems in Perseus and in other galaxy clusters. In addition, deeper spectroscopic observations are needed to reconstruct the star formation histories of these dwarfs. This will be essential for distinguishing between different evolutionary pathways and for studying the GC systems associated with each population separately. At the same time, more complete observations should also be conducted for Virgo and Coma dwarfs and UDGs.

We have also demonstrated that large ground-based telescopes like Subaru can reliably measure the GC systems in LSB galaxies at the distance of Perseus cluster ($\sim$ 75 Mpc). This foreshadows the expected transformative contributions of the next generation of survey telescopes, such as the Vera C.\ Rubin Observatory and the Nancy Grace Roman Space Telescope, for GC studies in the nearby universe \citep{Usher2023,Dage2023,Romanowsky2025}.

\begin{acknowledgments}

We thank the referee for useful feedback. 
We thank Teymoor Saifollahi for early data sharing of Perseus dwarfs from Euclid, Alfonso Aguerri for sharing their redshift catalog, and Laura Sales for helpful discussion.
We thank Joel Pfeffer for sharing data from E-MOSAICS. This work is supported by National Science Foundation grant AST-2308390 and by NASA under award No.~80NSSC25K7741. A.F.M. has received support from RYC2021-031099-I and PID2021-123313NA-I00 of MICIN/AEI/10.13039/501100011033/FEDER,UE, NextGenerationEU/PRT. D.A.F. thanks ARC for financial support via DP220101863. 
This work was supported by a NASA Keck PI Data Award, administered by the NASA Exoplanet Science Institute. Data presented herein were obtained at the W. M. Keck Observatory from telescope time allocated to the National Aeronautics and Space Administration through the agency's scientific partnership with the California Institute of Technology and the University of California.
The Observatory was made possible by the generous financial support of the W. M. Keck Foundation. The authors wish to recognize and acknowledge the very significant cultural role and reverence that the summit of Maunakea has always had within the indigenous Hawaiian community. We are most fortunate to have the opportunity to conduct observations from this mountain.
Research reported in this publication was supported by the Division of Research and Innovation at San Jos\'e State University under Award Number 25-RSG-08-135. The content is solely the responsibility of the authors and does not necessarily represent the official views of San Jos\'e State University.

\end{acknowledgments}

\bibliography{reference}{}
\bibliographystyle{aasjournal}

\restartappendixnumbering
\appendix

\section{Dwarf gallery and properties}
\label{sec:dwarf_gallery_properties}

\begin{figure*}
\centering
\includegraphics[width=0.96\textwidth]{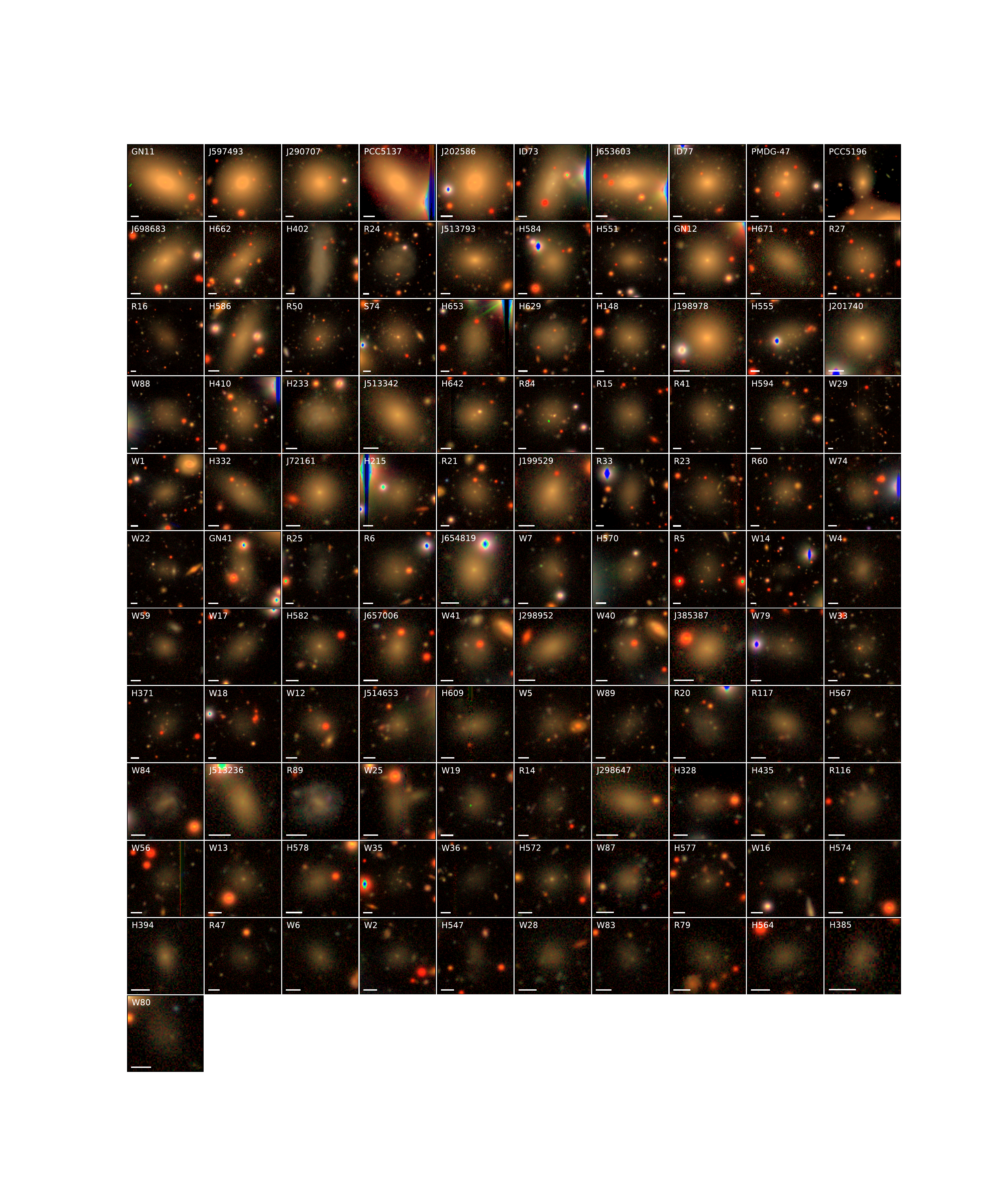}
\caption{The Subaru/HSC imaging gallery of the Perseus dwarfs in our sample, ordered by the HSC $g$-band magnitude from left to right and top to bottom (brightest first). The pseudocolor images are created using the HSC $g$, $r$, and $i$ bands, with the \cite{Lupton2004} algorithm. The cutout image size is approximately 6 times the circularized effective radius of each galaxy, and the white scale bars at the lower left of each panel indicate 1 kpc physical size. North is up and east is left.}
\label{fig:gallery}
\end{figure*}

\begin{figure*}
\centering
\includegraphics[width=0.96\textwidth]{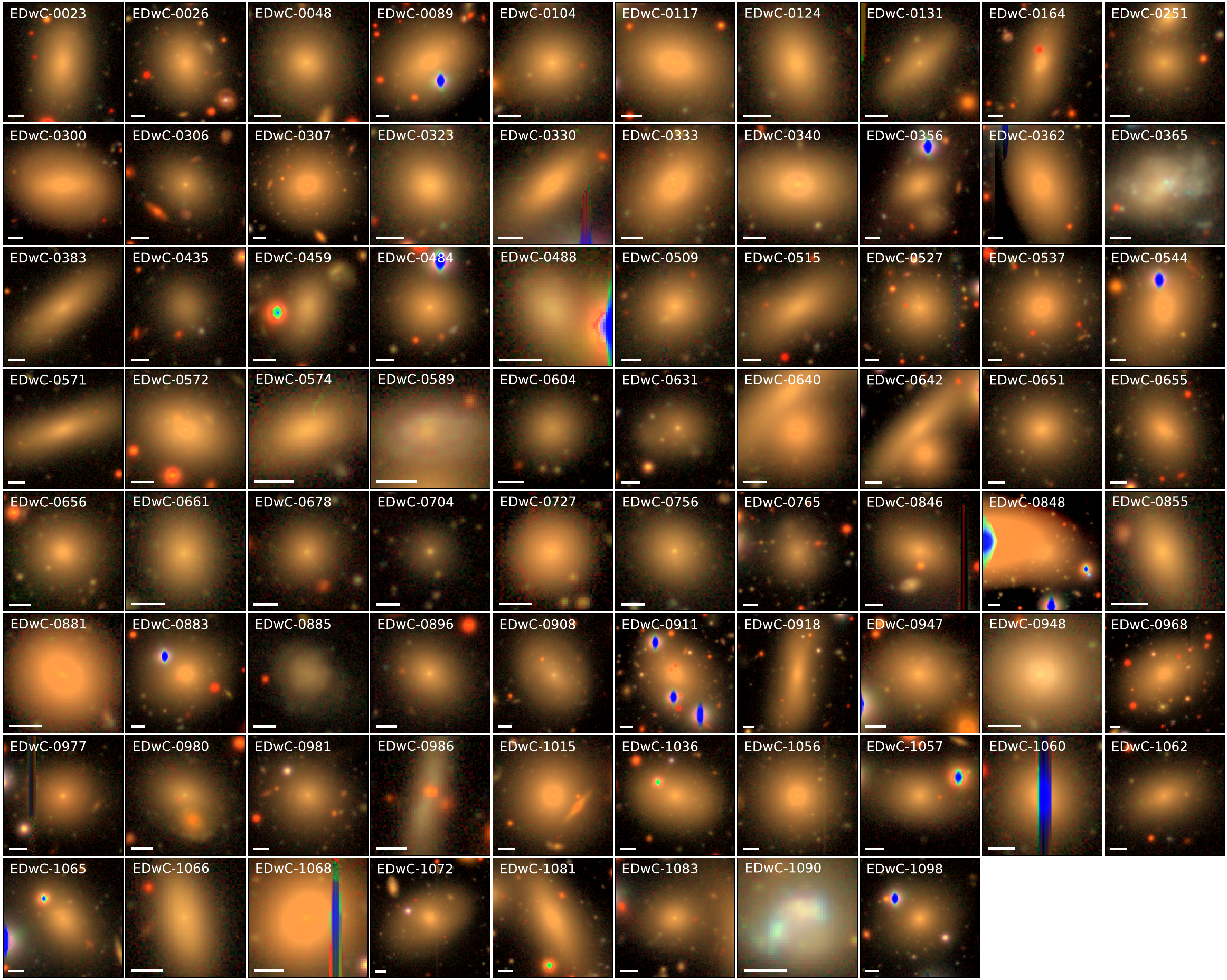}
\caption{Similar to Figure \ref{fig:gallery}, but for our supplementary sample from Euclid, ordered by their IDs in \cite{Marleau2025}, from left to right and top to bottom.}
\label{fig:gallery_euclid}
\end{figure*}

We present Subaru HSC color images of the dwarfs in our sample with Keck or HST observations in Figure \ref{fig:gallery}, with their properties listed in Table \ref{tab:data_properties}. The supplementary subsample from Euclid has the Subaru HSC color images shown in Figure \ref{fig:gallery_euclid}, and their properties are listed in Table \ref{tab:data_properties_euclid}. In addition to the identifiers used in this work, many dwarfs are also cataloged under different names in the literature. Table \ref{tab:data_properties_2} compiles all these identifiers for reference.

\begin{figure}
\centering
\includegraphics[width=0.472\textwidth]{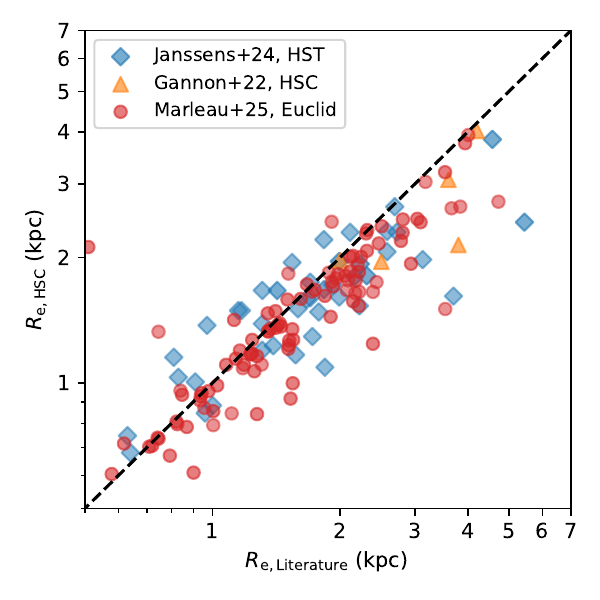}
\caption{Comparisons between our $R_{\rm e}$ measurements and literature values for the same dwarfs.}
\label{fig:size_literature}
\end{figure}

As mentioned in Section \ref{sec:galfit}, the Subaru HSC imaging data used in this work have the issue of background oversubtraction. To further assess the impact, we compare our size measurements with those reported in the literature for the same dwarfs, as shown in Figure \ref{fig:size_literature}. The literature values of $R_{\rm e}$ are based on different filters: HST F814W in \cite{Janssens2024}, HSC $g$-band in \cite{Gannon2022}, and Euclid $I_{\rm E}$ in \cite{Marleau2025}. To ensure a fair comparison, we use our HSC size measurements in filters with similar wavelengths -- specifically, $i$, $g$, and the average of $r$ and $i$, respectively. 

Overall, our measurements show good agreement with the literature. For dwarfs with $R_{\rm e} \gtrsim 2$ kpc, our sizes tend to be slightly smaller, likely due to the oversubtraction issue, although the discrepancy remains within an acceptable range. Other outliers are generally due to the presence of nearby giant galaxies, which makes background subtraction more challenging and uncertain. The largest scatter is observed between our measurements and those from HST. However, a similarly large scatter is also seen between HST and Euclid measurements, suggesting that HST may not be optimal for morphological measurements of LSB galaxies. %For properly calibrated images, 
The consistency of $R_{\rm e}$ measurements also implies that derived magnitudes and other morphological parameters are similarly reliable.

\clearpage

\startlongtable
\begin{deluxetable*}{lcccccccccc}
\tabletypesize{\footnotesize}
\renewcommand{\arraystretch}{0.88}
\tablewidth{0pt}
\tablecaption{Properties of the dwarf galaxies in our sample, sorted by $g$-band magnitude.
\label{tab:data_properties}}
\tablehead{
\colhead{Galaxy} & \colhead{R.A.} & \colhead{Decl.} & \colhead{$V_r$} & \colhead{$R_{\rm e}$} & \colhead{$b/a$} & \colhead{$m_g$} & \colhead{$(g-r)_0$} & \colhead{$(g-i)_0$} & \colhead{$\langle \mu_{g} \rangle_{\rm e}$} & \colhead{$S_M$} \\
\colhead{} & \colhead{(deg)} & \colhead{(deg)} & \colhead{(km s${}^{-1}$)} & \colhead{(kpc)} & \colhead{} & \colhead{(mag)} & \colhead{(mag)} & \colhead{(mag)} & \colhead{(${\rm mag/arcsec^2}$)} & \colhead{}}
\startdata
    GN11 & 50.14141 & 41.22190 & 6753 & 2.44 & 0.59 & 16.48 & 0.70 & 1.05 & 22.04 & -- \\
    J597493 & 50.57021 & 41.55253 & 6013 & 1.85 & 0.87 & 16.89 & 0.68 & 1.02 & 22.27 & -- \\
    J290707 & 50.82523 & 41.56793 & 6033 & 2.00 & 0.96 & 17.13 & 0.63 & 0.93 & 22.78 & $0.03 \pm 0.03$ \\
    PCC5137 & 49.96351 & 41.30973 & 6273 & 1.61 & 0.61 & 17.29 & 0.63 & 1.05 & 21.99 & $-0.02 \pm 0.03$${}^e$ \\
    J202586 & 50.11440 & 41.26932 & 4455 & 1.32 & 0.84 & 17.50 & 0.67 & 1.06 & 22.11 & -- \\
    ID73 & 49.48989 & 41.77581 & 7343 & 2.57 & 0.45 & 17.52 & 0.52 & 0.78 & 22.89 & $0.12 \pm 0.09$${}^e$ \\
    J653603 & 49.38538 & 41.82210 & 4061 & 1.68 & 0.54 & 17.68 & 0.60 & 0.95 & 22.32 & $0.53 \pm 0.10$${}^e$ \\
    ID77 & 49.60991 & 41.81876 & 5452 & 1.75 & 0.91 & 17.74 & 0.62 & 0.96 & 23.05 & $0.73 \pm 0.10$${}^e$ \\
    PMDG-47 & 50.14163 & 41.37964 & 4591${}^*$ & 1.95 & 0.95 & 17.75 & 0.64 & 1.02 & 23.33 & $0.33 \pm 0.07$${}^e$ \\
    PCC5196 & 49.97549 & 41.30897 & 5696 & 2.86 & 0.63 & 18.01 & 0.58 & 0.95 & 23.98 & $0.18 \pm 0.16$${}^e$ \\
    J698683 & 50.12727 & 41.18051 & 5707 & 1.90 & 0.60 & 18.34 & 0.63 & 0.92 & 23.37 & $0.25 \pm 0.06$ \\
    H662 & 50.92158 & 41.57217 & 5593${}^*$ & 2.52 & 0.50 & 18.42 & 0.63 & 0.94 & 23.85 & $0.83 \pm 0.13$ \\ %DEIMOS 5543
    H402 & 50.10498 & 41.21636 & 6960 & 3.17 & 0.30 & 18.44 & 0.35 & 0.52 & 23.85 & $0.66 \pm 0.25$ \\
    R24 & 49.64893 & 41.80897 & 7784${}^*$ & 3.07 & 0.82 & 18.46 & 0.42 & 0.62 & 24.88 & $0.61 \pm 0.32$${}^e$ \\ %DEIMOS 7687
    J513793 & 49.63645 & 41.83107 & 7653 & 1.80 & 0.72 & 18.59 & 0.66 & 0.98 & 23.70 & $0.58 \pm 0.15$${}^e$ \\
    H584 & 50.61420 & 41.65169 & 3373 & 1.76 & 0.86 & 18.72 & 0.61 & 0.85 & 23.98 & $-0.15 \pm 0.05$ \\
    H551 & 50.53385 & 41.75051 & 5748 & 2.65 & 0.75 & 18.73 & 0.52 & 0.85 & 24.72 & $0.11 \pm 0.14$ \\
    GN12 & 50.22635 & 41.17183 & 4839 & 1.22 & 0.95 & 18.77 & 0.61 & 0.99 & 23.34 & $0.01 \pm 0.05$ \\
    H671 & 50.99235 & 41.49661 & 4995 & 1.87 & 0.65 & 18.88 & 0.62 & 0.88 & 23.97 & $0.30 \pm 0.15$ \\
    R27 & 49.93166 & 41.71304 & 4376${}^*$ & 1.88 & 0.92 & 18.90 & 0.64 & 0.96 & 24.37 & $1.28 \pm 0.20$${}^h$ \\
    R16 & 49.65195 & 41.19210 & 4679${}^*$ & 4.02 & 0.65 & 18.95 & 0.64 & 0.97 & 25.69 & $0.62 \pm 0.18$${}^h$ \\
    H586 & 50.61892 & 41.53714 & 6436${}^*$ & 2.03 & 0.43 & 18.96 & 0.55 & 0.80 & 23.77 & $0.02 \pm 0.07$ \\ %DEIMOS 6416
    R50 & 49.44760 & 41.79332 & 4048 & 2.64 & 0.85 & 19.03 & 0.53 & 0.85 & 25.16 & $2.39 \pm 0.45$${}^e$ \\
    S74 & 49.29830 & 41.16789 & 6215${}^*$ & 2.14 & 0.92 & 19.03 & 0.57 & 0.98 & 24.79 & $1.91 \pm 0.25$ \\
    H653 & 50.89854 & 41.56002 & 4943 & 2.30 & 0.58 & 19.06 & 0.57 & 0.83 & 24.47 & $0.54 \pm 0.23$ \\
    H629 & 50.74297 & 41.53384 & 5548${}^*$ & 1.72 & 0.84 & 19.08 & 0.46 & 0.68 & 24.26 & $-0.10 \pm 0.11$ \\ %DEIMOS 5613
    H148 & 49.50569 & 41.82684 & 5380 & 1.85 & 0.90 & 19.08 & 0.58 & 0.85 & 24.48 & $0.88 \pm 0.25$${}^e$ \\
    J198978 & 50.07025 & 41.32296 & 5626 & 0.88 & 0.95 & 19.10 & 0.67 & 1.01 & 22.97 & $0.24 \pm 0.08$ \\
    H555 & 50.54675 & 41.59346 & 2866 & 2.28 & 0.57 & 19.13 & 0.50 & 0.74 & 24.51 & $1.20 \pm 0.43$ \\
    J201740 & 50.08133 & 41.28206 & 4805 & 0.92 & 0.97 & 19.19 & 0.55 & 0.79 & 23.16 & $0.69 \pm 0.22$ \\
    W88 & 49.99624 & 41.30915 & 6767${}^*$ & 2.64 & 0.73 & 19.21 & 0.46 & 0.75 & 25.18 & $1.72 \pm 0.63$${}^h$ \\ %DEIMOS 6710
    H410 & 50.13390 & 41.19850 & 5209${}^*$ & 1.88 & 0.82 & 19.28 & 0.64 & 0.96 & 24.63 & $1.55 \pm 0.31$ \\ %DEIMOS 5202
    H233 & 49.72326 & 41.79780 & 7908 & 1.36 & 0.87 & 19.29 & 0.60 & 0.81 & 23.99 & $0.47 \pm 0.44$${}^e$ \\
    J513342 & 49.60240 & 41.84327 & 7482 & 1.10 & 0.70 & 19.34 & 0.65 & 0.95 & 23.35 & $0.09 \pm 0.09$${}^e$ \\
    H642 & 50.82242 & 41.59516 & 4740 & 1.49 & 0.88 & 19.35 & 0.56 & 0.81 & 24.26 & $-0.17 \pm 0.10$ \\
    R84 & 49.35379 & 41.73932 & 4039${}^*$ & 1.97 & 0.95 & 19.38 & 0.53 & 0.88 & 24.98 & $2.37 \pm 0.45$${}^h$ \\
    R15 & 49.26581 & 41.24864 & 4762${}^*$ & 1.95 & 0.94 & 19.39 & 0.52 & 0.76 & 24.96 & $0.66 \pm 0.26$${}^h$ \\
    R41 & 50.21793 & 41.15645 & 6958 & 1.93 & 0.92 & 19.40 & 0.59 & 0.83 & 24.94 & $1.08 \pm 0.19$ \\
    H594 & 50.63152 & 41.56464 & 5169 & 1.45 & 0.97 & 19.43 & 0.59 & 0.84 & 24.40 & $0.51 \pm 0.23$ \\
    W29 & 49.59721 & 41.75009 & -- & 4.04 & 0.70 & 19.44 & 0.40 & 0.62 & 26.28 & $1.23 \pm 0.71$${}^h$ \\
    W1 & 49.25156 & 41.32243 & 4614${}^*$ & 2.33 & 0.80 & 19.45 & 0.39 & 0.57 & 25.23 & $0.69 \pm 0.62$${}^h$ \\
    H332 & 49.95743 & 41.66925 & 3196${}^*$ & 1.86 & 0.52 & 19.48 & 0.59 & 0.82 & 24.29 & $0.15 \pm 0.25$${}^e$ \\
    J72161 & 50.58676 & 41.76617 & 4721 & 1.00 & 0.94 & 19.53 & 0.64 & 1.09 & 23.65 & $0.00 \pm 0.09$ \\
    H215 & 49.66779 & 41.82491 & 3325 & 1.57 & 0.90 & 19.57 & 0.59 & 0.92 & 24.63 & $0.37 \pm 0.32$${}^e$ \\
    R21 & 50.12271 & 41.74750 & 5589${}^*$ & 1.81 & 0.90 & 19.58 & 0.54 & 0.92 & 24.94 & $2.66 \pm 0.65$${}^h$ \\
    J199529 & 50.05385 & 41.30792 & 5210 & 0.98 & 0.75 & 19.62 & 0.61 & 0.86 & 23.45 & $0.30 \pm 0.11$ \\
    R33 & 50.10747 & 41.52000 & -- & 2.45 & 0.56 & 19.63 & 0.52 & 0.74 & 25.13 & $-0.43 \pm 0.28$${}^e$ \\
    R23 & 49.96450 & 41.90968 & -- & 2.11 & 0.88 & 19.65 & 0.68 & 0.97 & 25.33 & $0.32 \pm 0.19$${}^h$ \\
    R60 & 49.90094 & 41.95724 & 3568${}^*$ & 1.76 & 0.96 & 19.78 & 0.51 & 0.83 & 25.15 & $2.06 \pm 0.60$${}^h$ \\
    W74 & 49.84137 & 41.45622 & -- & 1.77 & 0.96 & 19.80 & 0.51 & 0.77 & 25.19 & $0.64 \pm 0.40$${}^e$ \\
    W22 & 49.52268 & 41.46180 & -- & 3.24 & 0.52 & 19.84 & 0.33 & 0.50 & 25.87 & $0.46 \pm 0.94$${}^h$ \\
    GN41 & 49.53189 & 41.79477 & 5198 & 2.02 & 0.51 & 19.85 & 0.49 & 0.78 & 24.84 & $3.59 \pm 0.70$${}^e$ \\
    R25 & 49.60944 & 41.69768 & -- & 2.67 & 0.54 & 19.90 & 0.31 & 0.43 & 25.56 & $-1.65 \pm 2.27$${}^h$ \\
    R6 & 49.24556 & 41.70292 & -- & 1.58 & 0.91 & 19.95 & 0.56 & 0.80 & 25.03 & $0.66 \pm 0.09$${}^h$ \\
    J654819 & 49.51757 & 41.80558 & 7299 & 0.84 & 0.81 & 20.01 & 0.53 & 0.78 & 23.58 & $0.13 \pm 0.33$${}^e$ \\
    W7 & 49.31673 & 41.33663 & -- & 1.59 & 0.88 & 20.09 & 0.55 & 0.77 & 25.15 & $0.28 \pm 0.64$${}^h$ \\
    H570 & 50.57598 & 41.51775 & 4913 & 1.64 & 0.76 & 20.11 & 0.44 & 0.68 & 25.08 & $2.49 \pm 1.60$ \\
    R5 & 49.39423 & 41.75610 & -- & 2.24 & 0.79 & 20.18 & 0.67 & 1.03 & 25.87 & $1.16 \pm 0.50$${}^h$ \\
    W14 & 49.41365 & 41.51789 & -- & 3.27 & 0.78 & 20.19 & 0.40 & 0.73 & 26.68 & $-0.20 \pm 0.79$${}^h$ \\
    W4 & 49.27966 & 41.38130 & -- & 1.61 & 0.82 & 20.20 & 0.53 & 0.84 & 25.21 & $0.45 \pm 0.88$${}^h$ \\
    W59 & 49.72628 & 41.25813 & -- & 1.59 & 0.83 & 20.27 & 0.44 & 0.65 & 25.26 & $0.19 \pm 0.89$${}^h$ \\
    W17 & 49.43382 & 41.35519 & -- & 1.80 & 0.60 & 20.29 & 0.54 & 0.80 & 25.21 & $0.75 \pm 0.49$${}^h$ \\
    H582 & 50.60752 & 41.62687 & 5615 & 1.14 & 0.97 & 20.36 & 0.48 & 0.71 & 24.79 & $-0.41 \pm 0.31$ \\
    J657006 & 49.50069 & 41.77071 & 7988 & 1.07 & 0.77 & 20.36 & 0.63 & 0.94 & 24.41 & $-0.10 \pm 0.21$${}^e$ \\
    W41 & 49.63867 & 41.68222 & -- & 1.15 & 0.96 & 20.38 & 0.51 & 0.76 & 24.84 & $-0.16 \pm 1.94$${}^h$ \\
    J298952 & 50.74864 & 41.60278 & 5808 & 1.02 & 0.64 & 20.40 & 0.53 & 0.80 & 24.15 & $0.66 \pm 0.29$ \\
    W40 & 49.63863 & 41.68215 & -- & 1.31 & 0.91 & 20.40 & 0.52 & 0.78 & 25.06 & $2.07 \pm 1.59$${}^h$ \\
    J385387 & 50.78256 & 41.54006 & 3767 & 0.72 & 0.86 & 20.45 & 0.58 & 0.96 & 23.76 & $-0.23 \pm 0.24$ \\
    W79 & 49.91327 & 41.20163 & -- & 1.60 & 0.75 & 20.46 & 0.60 & 0.75 & 25.36 & $0.90 \pm 0.64$${}^h$ \\
    W33 & 49.60811 & 41.68542 & -- & 1.67 & 0.90 & 20.48 & 0.57 & 0.77 & 25.66 & $4.06 \pm 1.74$${}^h$ \\
    H371 & 50.05061 & 41.27665 & 7432 & 2.00 & 0.82 & 20.50 & 0.60 & 0.80 & 25.98 & $0.54 \pm 0.34$ \\
    W18 & 49.45142 & 41.31099 & -- & 2.02 & 0.86 & 20.50 & 0.54 & 0.84 & 26.05 & $-0.16 \pm 0.80$${}^h$ \\
    W12 & 49.40295 & 41.38373 & -- & 1.41 & 0.81 & 20.56 & 0.65 & 0.89 & 25.26 & $0.54 \pm 0.55$${}^h$ \\
    J514653 & 49.67754 & 41.82493 & 7960 & 1.28 & 0.81 & 20.58 & 0.69 & 0.88 & 25.08 & $0.97 \pm 0.31$${}^e$ \\
    H609 & 50.68645 & 41.58487 & 3687 & 1.21 & 0.73 & 20.64 & 0.59 & 0.91 & 24.91 & $1.70 \pm 0.64$ \\
    W5 & 49.29590 & 41.56763 & -- & 1.52 & 0.84 & 20.69 & 0.67 & 1.00 & 25.59 & $0.81 \pm 0.56$${}^h$ \\
    W89 & 50.00055 & 41.28493 & 7048 & 1.84 & 0.66 & 20.79 & 0.47 & 0.65 & 25.86 & $4.88 \pm 2.47$${}^h$ \\
    R20 & 50.10254 & 41.72456 & -- & 1.27 & 0.81 & 20.80 & 0.46 & 0.66 & 25.29 & $0.27 \pm 1.26$${}^h$ \\
    R117 & 49.51559 & 41.45244 & -- & 1.08 & 0.71 & 20.85 & 0.51 & 0.78 & 24.84 & $0.37 \pm 1.06$${}^h$ \\
    H567 & 50.57076 & 41.51393 & 4830 & 1.35 & 0.87 & 20.89 & 0.57 & 0.82 & 25.58 & $-0.76 \pm 0.37$ \\
    W84 & 49.95705 & 41.72852 & -- & 1.04 & 0.84 & 20.91 & 0.35 & 0.74 & 25.01 & $0.22 \pm 1.54$${}^h$ \\
    J513236 & 49.61843 & 41.84446 & 6099 & 0.84 & 0.55 & 20.92 & 0.63 & 0.88 & 24.08 & $-0.25 \pm 0.33$${}^e$ \\
    R89 & 50.05336 & 41.74940 & -- & 0.68 & 0.91 & 20.93 & 0.28 & 0.38 & 24.20 & $-0.28 \pm 2.03$${}^h$ \\
    W25 & 49.56435 & 41.47642 & -- & 1.28 & 0.53 & 20.97 & 0.58 & 0.92 & 25.01 & $0.43 \pm 0.72$${}^h$ \\
    W19 & 49.47153 & 41.32558 & -- & 1.20 & 0.87 & 20.99 & 0.43 & 0.71 & 25.44 & $-0.89 \pm 1.16$${}^h$ \\
    R14 & 49.27562 & 41.21774 & -- & 1.86 & 0.57 & 21.02 & 0.49 & 0.81 & 25.94 & $-0.15 \pm 0.84$${}^h$ \\
    J298647 & 50.73265 & 41.60391 & 7041 & 0.81 & 0.58 & 21.04 & 0.59 & 0.84 & 24.19 & $-0.32 \pm 0.39$ \\
    H328 & 49.95181 & 41.30861 & 6071 & 1.19 & 0.69 & 21.05 & 0.65 & 1.00 & 25.21 & $2.13 \pm 0.40$ \\
    H435 & 50.22173 & 41.13409 & 4725 & 1.11 & 0.81 & 21.10 & 0.60 & 0.88 & 25.29 & $0.13 \pm 0.40$ \\
    R116 & 49.44157 & 41.50329 & -- & 0.87 & 0.93 & 21.11 & 0.48 & 0.72 & 24.93 & $0.67 \pm 1.47$${}^h$ \\
    W56 & 49.70021 & 41.23392 & -- & 1.45 & 0.79 & 21.11 & 0.53 & 0.82 & 25.85 & $2.00 \pm 1.52$${}^h$ \\
    W13 & 49.40924 & 41.53251 & -- & 1.14 & 0.88 & 21.16 & 0.62 & 1.00 & 25.49 & $0.66 \pm 0.99$${}^h$ \\
    H578 & 50.59975 & 41.60681 & 4326 & 1.05 & 0.68 & 21.22 & 0.53 & 0.82 & 25.10 & $0.96 \pm 0.56$ \\
    W35 & 49.61753 & 41.66361 & -- & 1.70 & 0.87 & 21.25 & 0.57 & 0.94 & 26.45 & $0.56 \pm 3.10$${}^h$ \\
    W36 & 49.62164 & 41.69425 & -- & 1.76 & 0.72 & 21.29 & 0.46 & 0.57 & 26.36 & $-1.26 \pm 5.04$${}^h$ \\
    H572 & 50.58215 & 41.58435 & 6282 & 1.10 & 0.87 & 21.32 & 0.46 & 0.81 & 25.56 & $1.59 \pm 0.77$ \\
    W87 & 49.98915 & 41.49210 & -- & 0.85 & 0.81 & 21.33 & 0.47 & 0.75 & 24.94 & $-0.52 \pm 1.48$${}^e$ \\
    H577 & 50.59956 & 41.65591 & 5159 & 1.22 & 0.98 & 21.40 & 0.50 & 0.80 & 26.00 & $2.33 \pm 0.73$ \\
    W16 & 49.42410 & 41.40063 & -- & 1.50 & 0.63 & 21.41 & 0.53 & 0.93 & 25.99 & $-0.37 \pm 1.35$${}^h$ \\
    H574 & 50.59769 & 41.71225 & 5347 & 1.44 & 0.45 & 21.44 & 0.64 & 0.72 & 25.54 & $1.13 \pm 0.47$ \\
    H394 & 50.09683 & 41.26279 & 4888 & 0.94 & 0.61 & 21.46 & 0.47 & 0.65 & 24.99 & $-0.59 \pm 0.84$ \\
    R47 & 50.05889 & 41.26173 & 5508 & 1.37 & 0.88 & 21.48 & 0.41 & 0.64 & 26.22 & $14.43 \pm 4.11$ \\
    W6 & 49.30532 & 41.36882 & -- & 1.12 & 0.73 & 21.53 & 0.62 & 0.70 & 25.64 & $0.00 \pm 1.61$${}^h$ \\
    W2 & 49.26362 & 41.34142 & -- & 1.17 & 0.75 & 21.62 & 0.63 & 1.17 & 25.84 & $0.16 \pm 1.82$${}^h$ \\
    H547 & 50.52469 & 41.73581 & 5700 & 1.55 & 0.49 & 21.71 & 0.53 & 0.83 & 26.09 & $3.05 \pm 1.67$ \\
    W28 & 49.59019 & 41.75764 & -- & 0.80 & 0.89 & 21.91 & 0.64 & 0.75 & 25.49 & $0.40 \pm 1.91$${}^h$ \\
    W83 & 49.94757 & 41.73595 & -- & 1.01 & 0.77 & 21.99 & 0.36 & 0.57 & 25.92 & $3.74 \pm 4.06$${}^h$ \\
    R79 & 49.58858 & 41.77117 & -- & 0.95 & 0.73 & 22.00 & 0.82 & 1.08 & 25.75 & $-0.35 \pm 1.16$${}^h$ \\
    H564 & 50.56749 & 41.70316 & 6376 & 0.81 & 0.76 & 22.05 & 0.60 & 0.82 & 25.49 & $3.36 \pm 1.42$ \\
    H385 & 50.08130 & 41.26505 & 6917 & 0.60 & 0.69 & 22.41 & 0.55 & 0.85 & 25.09 & $0.23 \pm 1.45$ \\
    W80 & 49.91327 & 41.22870 & -- & 0.79 & 0.74 & 22.55 & 0.60 & 0.73 & 25.91 & $3.52 \pm 3.41$${}^h$ \\
\enddata
\tablecomments{The radial velocity $V_r$ is measured from Keck/DEIMOS, but the Keck/KCWI data are preferentially used if available (marked with ``*"). All photometric properties are obtained from GALFIT, including major-axis effective radius $R_{\rm e}$ and axis ratio $b/a$ in the $g$-band, and extinction-corrected $g$-band magnitude, $g-r$ and $g-i$ colors, and mean $g$-band surface brightness within the effective radius. The last column shows the GC system specific mass (percent mass fraction) for each galaxy, marked with ``$h$" if measured from HST, and ``$e$" if measured from Euclid, otherwise from HSC.}
\end{deluxetable*}

\startlongtable
\begin{deluxetable*}{lcccccccccc}
\tabletypesize{\footnotesize}
\renewcommand{\arraystretch}{0.88}
\tablewidth{0pt}
\tablecaption{Properties of the Euclid dwarf galaxies in our sample not included in Table \ref{tab:data_properties}, sorted by their Euclid ID.
\label{tab:data_properties_euclid}}
\tablehead{
\colhead{Galaxy} & \colhead{R.A.} & \colhead{Decl.} & \colhead{$V_r$} & \colhead{$R_{\rm e}$} & \colhead{$b/a$} & \colhead{$m_g$} & \colhead{$(g-r)_0$} & \colhead{$(g-i)_0$} & \colhead{$\langle \mu_{g} \rangle_{\rm e}$} & \colhead{$S_M$} \\
\colhead{} & \colhead{(deg)} & \colhead{(deg)} & \colhead{(km s${}^{-1}$)} & \colhead{(kpc)} & \colhead{} & \colhead{(mag)} & \colhead{(mag)} & \colhead{(mag)} & \colhead{(${\rm mag/arcsec^2}$)} & \colhead{}}
\startdata
EDwC-0023 & 49.12958 & 41.69727 & 4635 & 2.05 & 0.54 & 17.79 & 0.62 & 0.92 & 22.87 & $0.40 \pm 0.08$ \\
EDwC-0026 & 49.14867 & 41.60580 & 3727 & 1.86 & 0.81 & 17.67 & 0.63 & 0.98 & 22.97 & $0.81 \pm 0.09$ \\
EDwC-0048 & 49.22780 & 41.61143 & 6936 & 0.84 & 0.91 & 19.11 & 0.61 & 0.90 & 22.81 & $-0.04 \pm 0.10$ \\
EDwC-0089 & 49.30482 & 41.33334 & 5878 & 2.25 & 0.69 & 16.76 & 0.64 & 0.99 & 22.31 & $0.13 \pm 0.05$ \\
EDwC-0104 & 49.32561 & 41.49292 & 5474 & 1.01 & 0.94 & 18.55 & 0.60 & 0.92 & 22.71 & $1.00 \pm 0.16$ \\
EDwC-0117 & 49.34765 & 41.47005 & 5814 & 1.26 & 0.69 & 17.48 & 0.62 & 0.92 & 21.78 & $-0.08 \pm 0.03$ \\
EDwC-0124 & 49.35520 & 41.82390 & 6648 & 0.97 & 0.67 & 18.71 & 0.62 & 0.92 & 22.41 & $-0.02 \pm 0.06$ \\
EDwC-0131 & 49.36627 & 41.35350 & 6609 & 1.36 & 0.54 & 19.16 & 0.61 & 0.87 & 23.34 & $0.30 \pm 0.15$ \\
EDwC-0164 & 49.39360 & 41.62140 & 4872 & 2.41 & 0.44 & 17.84 & 0.67 & 0.99 & 23.05 & $0.03 \pm 0.05$ \\
EDwC-0251 & 49.47309 & 41.85749 & 4392 & 1.17 & 0.93 & 18.88 & 0.65 & 0.96 & 23.35 & $0.13 \pm 0.11$ \\
EDwC-0300 & 49.51840 & 41.36970 & 5295 & 2.05 & 0.62 & 17.16 & 0.68 & 1.05 & 22.38 & $0.15 \pm 0.03$ \\
EDwC-0306 & 49.52312 & 41.43548 & 5555 & 1.31 & 0.90 & 19.03 & 0.59 & 0.94 & 23.70 & $0.37 \pm 0.15$ \\
EDwC-0307 & 49.52442 & 41.58136 & 5754 & 2.07 & 0.89 & 16.66 & 0.67 & 1.04 & 22.30 & $0.60 \pm 0.06$ \\
EDwC-0323 & 49.54182 & 42.05774 & 5003 & 0.83 & 0.88 & 18.70 & 0.58 & 0.87 & 22.33 & $0.08 \pm 0.08$ \\
EDwC-0330 & 49.54858 & 41.61685 & 3848 & 1.31 & 0.46 & 18.36 & 0.63 & 0.93 & 22.30 & $0.11 \pm 0.07$ \\
EDwC-0333 & 49.55017 & 41.42334 & 5158 & 1.20 & 0.69 & 17.70 & 0.63 & 0.95 & 21.89 & $0.32 \pm 0.07$ \\
EDwC-0340 & 49.55458 & 41.80251 & 4835 & 1.21 & 0.67 & 17.08 & 0.56 & 0.84 & 21.26 & $-0.01 \pm 0.03$ \\
EDwC-0356 & 49.56666 & 41.27755 & 5078 & 1.71 & 0.82 & 18.02 & 0.53 & 0.90 & 23.16 & $0.06 \pm 0.09$ \\
EDwC-0362 & 49.57413 & 41.50309 & 5738 & 1.91 & 0.62 & 17.20 & 0.68 & 1.04 & 22.27 & $0.29 \pm 0.05$ \\
EDwC-0365 & 49.57605 & 41.48655 & 7605 & 1.38 & 0.57 & 17.13 & 0.21 & 0.26 & 21.40 & $0.54 \pm 0.25$ \\
EDwC-0383 & 49.59598 & 41.19746 & 7376 & 2.25 & 0.41 & 18.13 & 0.60 & 0.87 & 23.12 & $-0.10 \pm 0.07$ \\
EDwC-0435 & 49.62927 & 41.63274 & 5156 & 1.32 & 0.87 & 19.60 & 0.57 & 0.88 & 24.24 & $0.28 \pm 0.24$ \\
EDwC-0459 & 49.64306 & 41.49250 & 3691 & 1.27 & 0.64 & 18.90 & 0.56 & 0.84 & 23.12 & $-0.31 \pm 0.12$ \\
EDwC-0484 & 49.66724 & 41.41730 & 6761 & 1.32 & 0.90 & 18.38 & 0.61 & 0.95 & 23.07 & $-0.21 \pm 0.08$ \\
EDwC-0488 & 49.66948 & 41.62422 & 6070 & 0.56 & 0.73 & 19.76 & 0.50 & 0.78 & 22.36 & $-0.38 \pm 0.25$ \\
EDwC-0509 & 49.68688 & 41.63382 & 4407 & 1.17 & 0.84 & 18.33 & 0.61 & 0.92 & 22.69 & $0.20 \pm 0.10$ \\
EDwC-0515 & 49.69250 & 41.40483 & 5376 & 1.76 & 0.49 & 18.24 & 0.68 & 0.98 & 22.88 & $0.00 \pm 0.08$ \\
EDwC-0527 & 49.69922 & 41.74782 & 4794 & 1.81 & 0.94 & 17.59 & 0.63 & 0.93 & 23.00 & $0.61 \pm 0.08$ \\
EDwC-0537 & 49.70765 & 41.50664 & 5061 & 1.65 & 0.95 & 17.27 & 0.64 & 1.01 & 22.50 & $0.37 \pm 0.05$ \\
EDwC-0544 & 49.71206 & 41.63980 & 4320 & 1.81 & 0.68 & 17.04 & 0.66 & 1.00 & 22.11 & $0.05 \pm 0.03$ \\
EDwC-0571 & 49.73340 & 41.57867 & 3309 & 2.33 & 0.34 & 18.19 & 0.62 & 0.92 & 23.04 & $-0.09 \pm 0.09$ \\
EDwC-0572 & 49.73342 & 41.68619 & 3478 & 1.16 & 0.74 & 17.87 & 0.61 & 0.91 & 22.05 & $0.12 \pm 0.06$ \\
EDwC-0574 & 49.73498 & 41.25990 & 5383 & 0.71 & 0.56 & 19.40 & 0.57 & 0.86 & 22.22 & $0.12 \pm 0.13$ \\
EDwC-0589 & 49.74353 & 42.04844 & 5658 & 0.78 & 0.48 & 19.41 & 0.42 & 0.71 & 22.26 & $0.30 \pm 0.28$ \\
EDwC-0604 & 49.75165 & 41.48401 & 5845 & 0.88 & 0.96 & 19.72 & 0.65 & 0.95 & 23.58 & $-0.08 \pm 0.14$ \\
EDwC-0631 & 49.77509 & 41.43855 & 6577 & 1.34 & 0.77 & 19.43 & 0.61 & 0.92 & 23.97 & $0.49 \pm 0.23$ \\
EDwC-0640 & 49.78228 & 41.41140 & 5090 & 0.94 & 0.95 & 18.42 & 0.64 & 1.04 & 22.41 & $-0.03 \pm 0.10$ \\
EDwC-0642 & 49.78263 & 41.41275 & 3906 & 2.48 & 0.31 & 17.75 & 0.61 & 0.93 & 22.65 & $0.15 \pm 0.15$ \\
EDwC-0651 & 49.78943 & 41.70467 & 4382 & 1.44 & 0.91 & 18.04 & 0.60 & 0.91 & 22.91 & $0.31 \pm 0.08$ \\
EDwC-0655 & 49.79257 & 41.67810 & 3066 & 1.62 & 0.76 & 18.07 & 0.63 & 0.96 & 23.01 & $0.38 \pm 0.10$ \\
EDwC-0656 & 49.79348 & 41.49350 & 7325 & 1.05 & 0.93 & 18.61 & 0.64 & 0.94 & 22.84 & $0.51 \pm 0.14$ \\
EDwC-0661 & 49.79950 & 41.75882 & 3493 & 0.72 & 0.80 & 19.50 & 0.63 & 0.85 & 22.72 & $0.04 \pm 0.12$ \\
EDwC-0678 & 49.81242 & 41.67776 & 4138 & 0.93 & 0.96 & 19.51 & 0.59 & 0.97 & 23.51 & $0.68 \pm 0.18$ \\
EDwC-0704 & 49.82825 & 41.44234 & 5110 & 0.93 & 0.95 & 20.75 & 0.50 & 0.76 & 24.74 & $0.72 \pm 0.68$ \\
EDwC-0727 & 49.85007 & 41.53315 & 4033 & 0.66 & 0.99 & 19.11 & 0.63 & 1.00 & 22.39 & $-0.43 \pm 0.18$ \\
EDwC-0756 & 49.87158 & 41.73608 & 7081 & 0.96 & 0.89 & 19.07 & 0.56 & 0.84 & 23.04 & $0.70 \pm 0.15$ \\
EDwC-0765 & 49.88235 & 41.52239 & 3696 & 1.64 & 0.89 & 18.60 & 0.54 & 0.83 & 23.75 & $-0.49 \pm 0.17$ \\
EDwC-0846 & 49.95740 & 41.32988 & 4780 & 1.50 & 0.74 & 18.32 & 0.61 & 0.94 & 23.06 & $0.23 \pm 0.10$ \\
EDwC-0848 & 49.95885 & 41.57278 & 3949 & 2.15 & 0.81 & 16.92 & 0.65 & 1.07 & 22.54 & $0.00 \pm 0.02$ \\
EDwC-0855 & 49.96554 & 41.39226 & 4290 & 0.76 & 0.58 & 19.42 & 0.55 & 0.85 & 22.42 & $0.12 \pm 0.08$ \\
EDwC-0881 & 49.99320 & 41.54785 & 5903 & 0.72 & 0.84 & 18.23 & 0.69 & 1.11 & 21.52 & $-0.38 \pm 0.04$ \\
EDwC-0883 & 49.99472 & 41.75004 & 6389 & 1.84 & 0.89 & 17.28 & 0.63 & 0.94 & 22.67 & $0.16 \pm 0.04$ \\
EDwC-0885 & 49.99804 & 41.81072 & 6361 & 1.10 & 0.84 & 19.59 & 0.47 & 0.67 & 23.81 & $0.56 \pm 0.22$ \\
EDwC-0896 & 50.01655 & 41.33466 & 4650 & 1.20 & 0.83 & 18.83 & 0.57 & 0.88 & 23.21 & $0.34 \pm 0.12$ \\
EDwC-0908 & 50.02589 & 42.00442 & 3364 & 2.13 & 0.68 & 17.82 & 0.57 & 0.86 & 23.24 & $0.07 \pm 0.12$ \\
EDwC-0911 & 50.03040 & 41.53720 & 5311 & 2.31 & 0.77 & 17.05 & 0.63 & 1.04 & 22.77 & $0.28 \pm 0.06$ \\
EDwC-0918 & 50.03768 & 41.81722 & 6177 & 3.75 & 0.33 & 17.23 & 0.68 & 1.02 & 23.08 & $0.26 \pm 0.05$ \\
EDwC-0947 & 50.06155 & 41.54400 & 4825 & 1.13 & 0.87 & 18.42 & 0.70 & 1.03 & 22.72 & $0.13 \pm 0.13$ \\
EDwC-0948 & 50.06235 & 41.94850 & 4946 & 0.69 & 0.88 & 17.81 & 0.45 & 0.66 & 21.06 & $-0.01 \pm 0.06$ \\
EDwC-0968 & 50.07992 & 41.71188 & 6342 & 3.05 & 0.67 & 16.74 & 0.70 & 1.09 & 22.92 & $0.44 \pm 0.06$ \\
EDwC-0977 & 50.09033 & 41.52693 & 3918 & 1.37 & 0.86 & 18.93 & 0.68 & 1.08 & 23.64 & $0.36 \pm 0.14$ \\
EDwC-0980 & 50.09168 & 41.73327 & 5046 & 1.15 & 0.78 & 19.15 & 0.59 & 0.91 & 23.38 & $0.86 \pm 0.15$ \\
EDwC-0981 & 50.09333 & 41.34953 & 3616 & 1.64 & 0.85 & 18.30 & 0.54 & 0.82 & 23.39 & $0.44 \pm 0.28$ \\
EDwC-0986 & 50.09888 & 41.86252 & 7581 & 1.48 & 0.22 & 19.63 & 0.44 & 0.71 & 23.06 & $1.49 \pm 0.37$ \\
EDwC-1015 & 50.13053 & 41.39007 & 4551 & 1.44 & 0.93 & 17.35 & 0.63 & 0.99 & 22.26 & $0.18 \pm 0.04$ \\
EDwC-1036 & 50.16215 & 41.77278 & 4969 & 1.97 & 0.63 & 17.70 & 0.68 & 1.03 & 22.87 & $0.29 \pm 0.06$ \\
EDwC-1056 & 50.20647 & 41.63166 & 5997 & 1.51 & 0.98 & 17.33 & 0.64 & 0.98 & 22.39 & $0.37 \pm 0.05$ \\
EDwC-1057 & 50.21060 & 41.39275 & 6974 & 1.47 & 0.73 & 18.56 & 0.59 & 0.92 & 23.25 & $0.00 \pm 0.08$ \\
EDwC-1060 & 50.21275 & 41.56685 & 5662 & 0.85 & 0.87 & 18.65 & 0.64 & 0.91 & 22.35 & $0.28 \pm 0.11$ \\
EDwC-1062 & 50.21782 & 41.53445 & 7058 & 1.76 & 0.64 & 18.50 & 0.63 & 0.95 & 23.43 & $0.39 \pm 0.12$ \\
EDwC-1065 & 50.22107 & 41.49445 & 4461 & 1.77 & 0.71 & 17.94 & 0.57 & 0.85 & 23.01 & $0.09 \pm 0.08$ \\
EDwC-1066 & 50.23332 & 41.66214 & 3235 & 0.99 & 0.49 & 19.00 & 0.60 & 0.91 & 22.39 & $0.09 \pm 0.08$ \\
EDwC-1068 & 50.23575 & 41.43880 & 6027 & 0.79 & 0.85 & 17.34 & 0.73 & 1.16 & 20.83 & $-0.02 \pm 0.01$ \\
EDwC-1072 & 50.24497 & 41.60930 & 6717 & 2.55 & 0.86 & 17.15 & 0.63 & 0.96 & 23.21 & $0.38 \pm 0.05$ \\
EDwC-1081 & 50.27529 & 41.53680 & 6072 & 2.48 & 0.44 & 17.21 & 0.62 & 0.96 & 22.48 & $0.31 \pm 0.10$ \\
EDwC-1083 & 50.28083 & 41.53817 & 5684 & 1.41 & 0.85 & 18.55 & 0.51 & 0.81 & 23.32 & $0.82 \pm 0.38$ \\
EDwC-1090 & 50.30712 & 41.65388 & 6884 & 0.59 & 0.67 & 17.88 & 0.11 & 0.12 & 20.48 & $0.23 \pm 0.23$ \\
EDwC-1098 & 50.34478 & 41.50308 & 5092 & 1.85 & 0.98 & 17.90 & 0.60 & 0.90 & 23.41 & $0.74 \pm 0.08$ \\
\enddata
\tablecomments{The radial velocity $V_r$ is reported in \cite{Kang2024}. The details of the photometric properties are the same as Table \ref{tab:data_properties}. The last column shows the GC system specific mass (percent mass fraction) measured from Euclid.}
\end{deluxetable*}

\startlongtable
\begin{deluxetable*}{lccl}
\tabletypesize{\footnotesize}
\renewcommand{\arraystretch}{1.05}
\tablewidth{0pt}
\tablecaption{Other galaxy identifications in the literature.
\label{tab:data_properties_2}}
\tablehead{
\colhead{Galaxy} & \colhead{R.A.} & \colhead{Decl.} & \colhead{Other Identifications} \\
\colhead{} & \colhead{(deg)} & \colhead{(deg)} & \colhead{} }
\startdata
    W1 & 49.25156 & 41.32243 & H60, PCC~0040 \\
    W2 & 49.26362 & 41.34142 & H63, PCC~0093, EDwC-0065 \\
    W4 & 49.27966 & 41.38130 & H69, PCC~0190, EDwC-0073 \\
    W5 & 49.29590 & 41.56763 & H73, PCC~0290, EDwC-0082 \\
    W6 & 49.30532 & 41.36882 & H76, PCC~0352, EDwC-0090 \\
    W7 & 49.31673 & 41.33663 & H78, PCC~0419, EDwC-0097 \\
    W12 & 49.40295 & 41.38373 & EDwC-0169 \\
    W13 & 49.40924 & 41.53251 & H108, PCC~0965, EDwC-0177 \\
    W14 & 49.41365 & 41.51789 & H110, PCC~0985, EDwC-0181 \\
    W16 & 49.42410 & 41.40063 & H113, PCC~1047, EDwC-0193 \\
    W17 & 49.43382 & 41.35519 & H118, PCC~1113, EDwC-0208 \\
    W18 & 49.45142 & 41.31099 & H124, PCC~1218, EDwC-0221 \\
    W19 & 49.47153 & 41.32558 & H133, PCC~1339, EDwC-0248 \\
    W22 & 49.52268 & 41.46180 & H152, PCC~1682, EDwC-0305 \\
    W25 & 49.56435 & 41.47642 & PCC~1925, EDwC-0352 \\
    W28 & 49.59019 & 41.75764 & H177, PCC~0376, EDwC-0376 \\
    W29 & 49.59721 & 41.75009 & H178, PCC~2157, EDwC-0387 \\
    W33 & 49.60811 & 41.68542 & H181, PCC~2251, EDwC-0403 \\
    W35 & 49.61753 & 41.66361 & H185, PCC~2335, EDwC-0417 \\
    W36 & 49.62164 & 41.69425 & H188, PCC~2371, EDwC-0424 \\
    W40 & 49.63863 & 41.68215 & PCC~2539, EDwC-0452 \\
    W41 & 49.63867 & 41.68222 & PCC~2548, EDwC-0454 \\
    W56 & 49.70021 & 41.23392 & H225, PCC~3023, EDwC-0528 \\
    W59 & 49.72628 & 41.25813 & H235, PCC~3239, EDwC-0563 \\
    W74 & 49.84137 & 41.45622 & H289, PCC~4141, EDwC-0717 \\
    W79 & 49.91327 & 41.20163 & H318, PCC~4735 \\
    W80 & 49.91327 & 41.22870 & H317, PCC~4736 \\
    W83 & 49.94757 & 41.73595 & H327, PCC~5007, EDwC-0836 \\
    W84 & 49.95705 & 41.72852 & H330, PCC~5094, EDwC-0845 \\ % spiral structure?
    W87 & 49.98915 & 41.49210 & PCC~5305, EDwC-0878 \\
    W88 & 49.99624 & 41.30915 & H352, PCC~5374, EDwC-0884 \\
    W89 & 50.00055 & 41.28493 & H355, PCC~5402 \\
    R5 & 49.39423 & 41.75610 & H105, EDwC-0165 \\
    R6 & 49.24556 & 41.70292 & H56, EDwC-0055 \\
    R14 & 49.27562 & 41.21774 & H68 \\
    R15 & 49.26581 & 41.24864 & H64 \\
    R16 & 49.65195 & 41.19210 & H207, EDwC-0471 \\
    R20 & 50.10254 & 41.72456 & H400, EDwC-0993 \\
    R21 & 50.12271 & 41.74750 & H406, EDwC-1011 \\
    R23 & 49.96450 & 41.90968 & H338, EDwC-0854 \\
    R24 & 49.64893 & 41.80897 & H203, EDwC-0468 \\
    R25 & 49.60944 & 41.69768 & H182, PCC~2262, EDwC-0406 \\
    R27 & 49.93166 & 41.71304 & H675, PCC~4867, EDwC-0823 \\
    R33 & 50.10747 & 41.52000 & H404, EDwC-0997 \\
    R41 & 50.21793 & 41.15645 & H432 \\
    R47 & 50.05889 & 41.26173 & H373 \\
    R50 & 49.44760 & 41.79332 & H122, EDwC-0218 \\
    R60 & 49.90094 & 41.95724 & H314, EDwC-0791 \\
    R79 & 49.58858 & 41.77117 & H176, EDwC-0373 \\
    R84 & 49.35379 & 41.73932 & H89, EDwC-0120 \\
    R89 & 50.05336 & 41.74940 & EDwC-0938 \\
    R116 & 49.44157 & 41.50329 & PCC~1161, EDwC-0214 \\
    R117 & 49.51559 & 41.45244 & H150, PCC~1630, EDwC-0293 \\
    PCC5137 & 49.96351 & 41.30973 & EDwC-0853 \\
    PCC5196 & 49.97549 & 41.30897 & EDwC-0863 \\
    PMDG-47 & 50.14163 & 41.37964 & ID47, EDwC-1025, K1366 \\
    ID73 & 49.48989 & 41.77581 & EDwC-0270, K536 \\
    ID77 & 49.60991 & 41.81876 & EDwC-0408, K680 \\
    H148 & 49.50569 & 41.82684 & EDwC-0285, K556 \\
    H215 & 49.66779 & 41.82491 & EDwC-0485 \\
    H233 & 49.72326 & 41.79780 & EDwC-0556 \\
    H328 & 50.05889 & 41.26173 & PCC~5040 \\
    H332 & 49.95743 & 41.66925 & EDwC-0847 \\
    J72161 & 50.58676 & 41.76617 & SDSS J032220.85+414558.0 \\
    J198978 & 50.07025 & 41.32296 & EDwC-0957, K1268 \\
    J199529 & 50.05385 & 41.30792 & EDwC-0940 \\
    J201740 & 50.08133 & 41.28206 & K1285 \\
    J202586 & 50.11440 & 41.26932 & K1323 \\
    J290707 & 50.82523 & 41.56793 & K2014 \\
    J298647 & 50.73265 & 41.60391 & SDSS J032255.87+413613.8 \\
    J298952 & 50.74864 & 41.60278 & SDSS J032259.70+413610.0 \\
    J385387 & 50.78256 & 41.54006 & SDSS J032307.81+413224.0 \\
    J513236 & 49.61843 & 41.84446 & EDwC-0419 \\
    J513342 & 49.60240 & 41.84327 & EDwC-0394 \\
    J513793 & 49.63645 & 41.83107 & EDwC-0449 \\
    J514653 & 49.67754 & 41.82493 & EDwC-0498 \\
    J597493 & 50.57021 & 41.55253 & K1860 \\
    J653603 & 49.38538 & 41.82210 & EDwC-0156, K428 \\
    J654819 & 49.51757 & 41.80558 & EDwC-0297 \\
    J657006 & 49.50069 & 41.77071 & EDwC-0281 \\
    J698683 & 50.12727 & 41.82210 & K1343 \\
    GN11 & 50.14141 & 41.22190 & LEDA 2177592, K1365 \\
    GN12 & 50.22635 & 41.17183 & K1466 \\
    GN41 & 49.53189 & 41.79477 & EDwC-0316 \\
\enddata
\tablecomments{Galaxy nomenclature origins are as follows:
W for \citet{Wittmann2017},
R for A.~Romanowsky/CFHT,
H and ID for S.~Huang/HSC,
J for S.~Janssens/HSC,
PCC for Perseus Cluster Catalog of \citet{Wittmann2019};
EDwC for Euclid;
K for \citet{Kang2024} redshift compilation.}
\end{deluxetable*}

\clearpage

\section{Euclid} \label{sec:euclid}

With the dwarf sample in Perseus from Euclid ERO \citep{Marleau2025}, we can double check our results in Section \ref{sec:results}. However, the galaxy stellar mass estimates for these dwarfs are not publicly available, so here we adopt specific frequency in $I_{\rm E}$-band, $S_{N, I_{\rm E}}$, instead of specific mass. We note that the GC number used for this practice is the average of \cite{Marleau2025} and \cite{Saifollahi2025}, and has been calibrated to HST as mentioned in Section \ref{sec:gc_richness}. Then, we can conduct a linear fit between $S_{N, I_{\rm E}}$ and magnitude $M_{I_{\rm E}}$:
$$\log S_{N, I_{\rm E}}=0.23 \times \log M_{I_{\rm E}}+4.17$$
The magnitude-independent specific frequency, $\Delta S_{N, I_{\rm E}}$, is therefore defined as the residual of log $S_{N, I_{\rm E}}$ from this linear relation. Similarly, we classify the dwarfs with $\Delta S_{N, I_{\rm E}} > 0.15$~dex as GC-rich, and $\Delta S_{N, I_{\rm E}} < -0.15$~dex as GC-poor. The remaining dwarfs can be considered GC-intermediate or ambiguous.

\begin{figure*}
\centering
\includegraphics[width=1\textwidth]{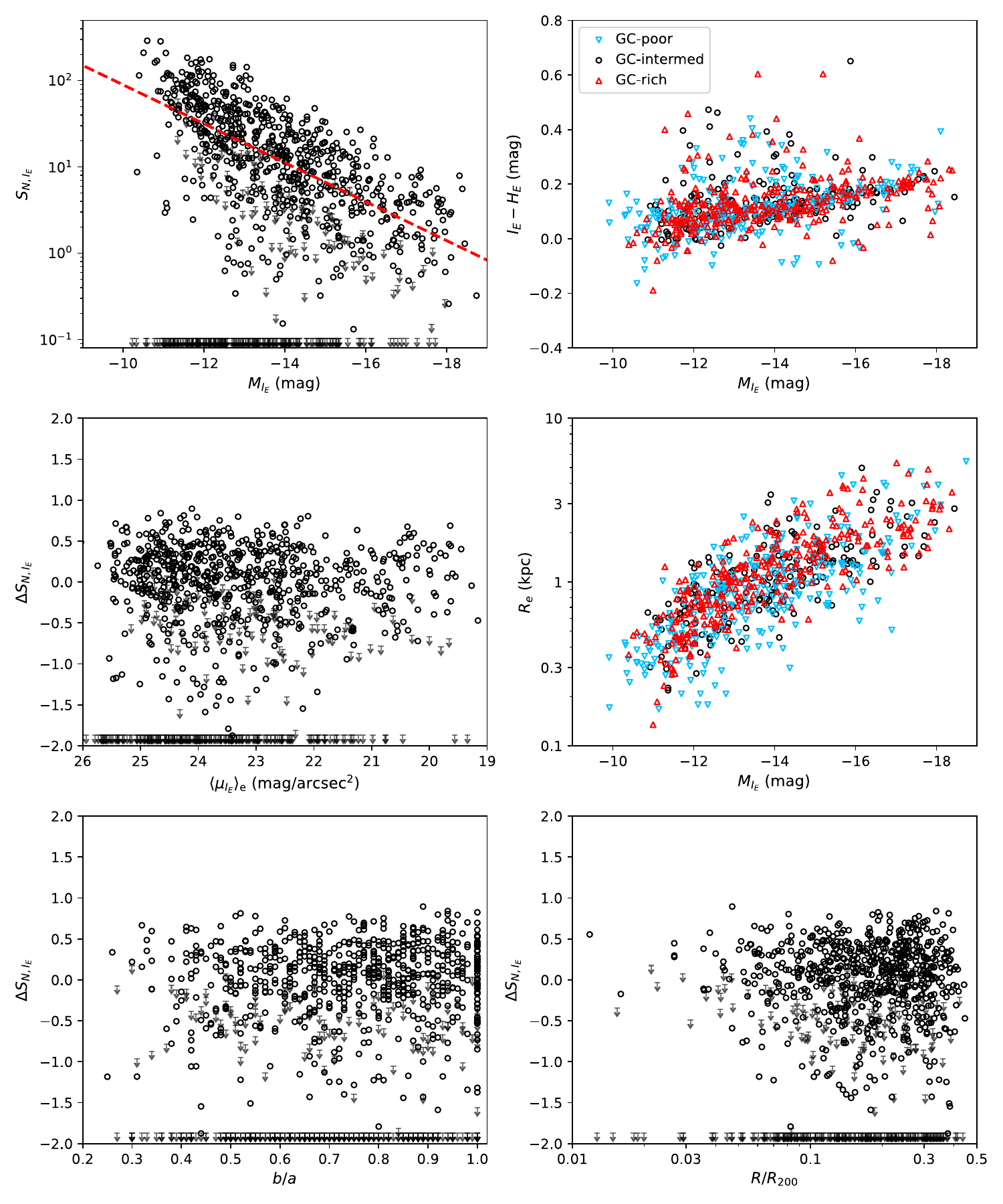}
\caption{Similar to Figures \ref{fig:mgc_mass}, \ref{fig:mgc_sb_ba}, \ref{fig:cmd_sizemag} and \ref{fig:mgc_infall}, but reproduced with the full Euclid dwarf galaxies sample in \cite{Marleau2025}. Specific frequency $S_{N,I_E}$ and magnitude-independent $\Delta S_{N,I_E}$ are used here instead of $S_M$ and $\Delta S_M$. In the first panel, the red dashed line represents the fit of $S_{N,I_E}$ with maximum likelihood linear regression. }
\label{fig:euclid_gc}
\end{figure*}

We have made similar versions of the figures in Section \ref{sec:results}, as shown in Figure \ref{fig:euclid_gc}. Compared to our results, the trends based on the full Euclid sample show no significant differences qualitatively. We can still observe correlations between GC richness and surface brightness or size. There remains a lack of correlation between GC richness and either axis ratio or clustercentric radius. Additionally, GC-poor dwarfs exhibit a larger color scatter compared to their GC-rich counterparts. However, we caution that up to 20\% of the dwarfs in this sample may not be true members of the Perseus cluster \citep{Marleau2025}. As a result, any intrinsic correlations could be diluted by foreground and background contamination, making them more difficult to detect.

\end{document}